\documentclass[a4paper,11pt]{article}
\pdfoutput=1 
\usepackage{jcappub} 
\usepackage[T1]{fontenc} 
\usepackage[utf8]{inputenc}
\usepackage{float}
\newcommand{\dd}{{\rm d}}

\newcommand{\sube}[1]{{\tiny \mbox{#1}}}

\title{\boldmath Constraining LTB models with JLA supernovae and BAO}

\author[a]{C. Z. Vargas}
\author[a]{F. T. Falciano}
\author[b,c]{and R. R. R. Reis}

\affiliation[a]{Centro Brasileiro de Pesquisas F\'\i sicas\\ Rua Xavier Sigaud st. 150, 22290-180, Rio de Janeiro, Brazil}
\affiliation[b]{Instituto de F\'\i sica, Universidade Federal do Rio de Janeiro, \\ Av. Athos da Silveira Ramos 149, 21941-972, Rio de Janeiro, RJ, Brazil}
\affiliation[c]{Observat\'orio do Valongo, Universidade Federal do Rio de Janeiro, \\ Ladeira do Pedro Ant\^onio 43, CEP 20080-090 Rio de Janeiro, RJ, Brazil}

\emailAdd{czuniga@cbpf.br}
\emailAdd{ftovar@cbpf.br}
\emailAdd{ribamar@if.ufrj.br}

\abstract{In the present work we constrain three different profiles of a Lema\^itre-Tolman-Bondi model using supernovae type Ia and baryon acoustic oscillation data. We improve common practice in the literature by carefully calibrating the supernovae in the appropriate inhomogeneous background dynamics. In addition, we address subtle issues in order to propagate the primordial BAO scale to present epoch. The combined analysis of BAO+SNIa offers a stringent test for these models. We use two distinct parameter estimation approaches, namely, the $\chi^2$ and the complete likelihood functional. It has been argued that these two approaches are not equivalent and indeed our analysis shows a specific example of their departure.
}

\begin{document}
\maketitle
\flushbottom

\section{Introduction}
\label{sec:Intro}

At the turn of the last century, observations of distant supernovae type Ia (SNIa)~\cite{Riess,Perlmutter} together with the observation of the cosmic microwave background~\cite{spergel2003} and the large scale structure~\cite{Tegmark2004,Seljak2005,Eisenstein2005,Semboloni2008} led us to consider that the universe could be accelerating. In the framework of a Friedmann-Lema\^itre-Robertson-Walker (FLRW) universe, the source for this acceleration has to be an exotic component dubbed dark energy. There are several theoretical candidates for dark energy such as quintessence fluids~\cite{ratra1988}, K-essence~\cite{Picon2001} and Chaplygin gas~\cite{Campos2013} just to name a few. There is also the possibility of mimicking dark energy by modifying the gravitational interaction. In these scenarios one changes General Relativity for other theories like $f(R)$ theories~\cite{Miranda2009}, DBI Galileons~\cite{Zumalacarregui2013} or Brane world cosmology~\cite{maier2011} that produce an early or late time accelerating expansion of the universe.

There is also a third approach which, within General Relativity (GR), gives a natural way to induce the present acceleration. Revoking the Copernican Principle (CP), one can construct inhomogeneous models that can suppress the need for dark energy~\cite{Wessel,Redlich,Mustapha,Celerier,Iguchi}. Their validity relies on the underdetermination of the model from observational data. Our data comes only from our past null cone and hence there is a collection of geometries compatible with the observational data. These models are effective with inhomogeneities of the order of a few fractions of the Hubble radius.

In a FLRW model the dynamics is entirely encoded in the scale factor. In inhomogeneous models, the spatial dependence gives extra degrees of freedom which in principle could better accommodate the data. Thus, it seems critical to have at least two independent observation to restrict inhomogeneous models. It is expedient that the number of free parameters of each model should be as lower as possible and the number of independent observational as high as possible. In particular, in this paper we use the Joint Light-curve Analysis sample~\cite{Betoule} Known as JLA sample. This extended sample of 740 supernovae combines low-redshift samples ($z<0.1$), the third year Sloan Digital Sky Survey sample (SDSS-II, $0.05<z<0.4$), the third year SuperNova Legacy Survey (SNLS, $0.2<z<1$) and the Hubble Space Telescope sample (HST, $z>1$).

The goal of this article is to improve and analyze the best fit of inhomogeneous models using SNIa and Baryon Acoustic Oscillation (BAO) data. We consider three different profiles of a Lema\^itre-Tolman-Bondi (LTB) model of the universe. Contrarily to some analysis in the literature, we carefully calibrate the SNIa data for a LTB dynamics. In a LTB universe, there are also subtle adjustments to propagate the primordial BAO scale to present epoch. The combined analysis of BAO+SNIa offers a stringent test for the models presented in this article. In addition, we test the validity of using the $\chi^2$ minimization as compared to the extremization of the likelihood functional. It has been argued that these two approaches are not equivalent and indeed our analysis shows a specific example of their departure.

The paper is organized as follows. In the next section we briefly review the main points of a LTB model and its dynamics. Section~\ref{sec:void_profiles} we specify the inhomogeneous matter profiles and in section~\ref{subsec:BAOinLTB} we describe how to incorporate the BAO physics in a LTB dynamics. In section~\ref{sec:obsdata} we characterize the SNIa and BAO data used to fit the models and discuss the results in section~\ref{sec:results}. Section~\ref{sec:Concl} is reserved for final comments and conclusions.

\section{The LTB model}\label{sec:LTB_model}
The standard cosmological model describes our universe being on average homogeneous and isotropic where the background is assumed to be a FLRW metric. This is the mathematical formulation of the Cosmological principle which is a possible way to implement the copernican principle, namely, the statement that we are not in a privileged location in the universe. For most of the last century, homogeneity and isotropy were only valuable assumptions but recently it has been shown they are consistent with observation (if one accepts the existence of dark energy and dark matter). Even though there is no direct confirmation of spatial homogeneity, observational data does support  spatial isotropy with respect to our point of observation~\cite{clarkson2012}. 

In the present work we shall drop homogeneity but still maintain the isotropy condition. Thus, the observable universe shall be modeled by an inhomogeneous but spherically symmetric metric which we assume to be of the LTB type~\cite{Lemaitre,Tolman,Bondi}. This family of metrics can be written\footnote{Units in which $c=1$ will be assumed in the following.} as
\begin{equation}\label{LTBmetric}
ds^2 = \dd t^2- \frac{X'^{2}(t,r)}{1+2E(r)} \dd r^2 - X^2(t,r)\dd \Omega ^2. 
\end{equation}
where $\dd \Omega ^2=\dd \theta^2+\sin^2\theta\,  \dd \varphi^2$ is the solid angle and prime denotes partial derivatives with respect to the radial coordinate, i.e~ $X'=\partial_rX$. The energy function $E(r)$ can be related to the non-constant spatial curvature and $X(t,r)$ plays the role of a scale function that can depends both on time and radial coordinates. It is straightforward to check that \eqref{LTBmetric} reduces to a FLRW metric when $X(t,r)=a(t)\, r$ and $2E(r)=-\kappa \, r^2$ with $\kappa=0,\pm1$.

The LTB metric \eqref{LTBmetric} has two effective scale factors, namely, the transverse or angular scale factor $X(t,r)$ which is associated with the area radius of each $S^2$ sphere and the parallel or radial scale factor $X'(r,t)$. Thus, contrasting to the FLRW metric which has only one Hubble factor, it is convenient to define two Hubble parameters
\begin{eqnarray}\label{Hsparam}
H_{\bot}(t,r) \equiv \frac{\dot{X}(t,r)}{X(t,r)} &\quad  \mbox{and} &\quad H_{\parallel}(t,r) \equiv \frac{\dot{X}^{\prime}(t,r)}{X^{\prime}(t,r)} \ ,
\end{eqnarray}
that represent respectively the transverse and radial expansion rates. The dots in the above equation denote partial derivatives with respect to time coordinates, i.e.~$\dot{X}=\partial_tX$. It is also possible to define a geometrical mean out of these Hubble parameters as 
\begin{equation}\label{HLTBmean}
\bar{H}_\sube{LTB}(t,r) = \Big[H_{\parallel}(t,r)\, H^2_{\bot}(t,r)\Big]^{1/3}\ .
\end{equation} 

The matter content is described by a pressureless inhomogeneous fluid whose stress-energy tensor can be written as $T^{\mu \nu} = \rho (t,r) u^{\mu} u^{\nu}$, where $u^{\mu}$ is the four-velocity field of the fluid. Due to the symmetries of \eqref{LTBmetric}, the analogue Friedmann-like equation reads
\begin{equation}\label{Xdot}
H_{\bot}^2 -\frac{2E}{X^2}= \frac{2M}{X^3} \ , 
\end{equation}
where $M(r)$ is another free function of $r$ that can be interpreted as the gravitational mass inside a spherical shell of radius $r$. This mass function is connected to the energy density through the relation
\begin{equation}\label{rhoM}
\rho(t,r)= \frac1{8\pi G}\frac{M'}{X^2X'} \ .
\end{equation}

Equations \eqref{Xdot} and \eqref{rhoM} compose the dynamic field equations for a dust LTB spacetime. The time derivative of \eqref{rhoM} combined with \eqref{Xdot} implies a continuity equation for the energy density
\begin{equation}\label{rhocont}
\dot{\rho}(t,r)+(2 H_{\bot}+H_{\parallel})\rho(t,r)=0\ .
\end{equation}

The above equation prompt us to define another mean Hubble factor through an arithmetic mean as
\begin{equation}\label{HLTBmean_arith}
\langle{H}\rangle (t,r)= \frac13\left(2 H_{\bot}+H_{\parallel}\right) \ .
\end{equation} 

The analogy between a LTB model and a FLRW universe can be carried further by defining a dimensionless matter and a curvature density parameters evaluated today given respectively as
\begin{eqnarray}
\Omega_M(r)= \frac{2M(r)}{H^2_{\bot 0}(r)X^3_0(r)}&\quad \mbox{and}&\quad \Omega_K(r) = \frac{2E(r)}{H^2_{\bot 0}(r)X^2_0(r)}\quad .
\end{eqnarray}

Note that equation \eqref{Xdot} means that these two density parameters are related by $\Omega _M(r) + \Omega _K(r) =1$. Additionally, labeling the present values of the Hubble factor and the scale function respectively as $H_{\bot 0}(r) \equiv H_{\bot}(t_0,r)$ and $X(t_0,r) \equiv X_0(r)$, we can recast \eqref{Xdot} as~\cite{EnqvistMattson,Enqvist}
\begin{equation} \label{Hubble}
H^2_{\bot}(t,r) = H^2_{\bot 0}(r) \left[ \Omega _M(r) \left(\frac{X_0(r)}{X(r,t)} \right)^3 + \Omega _K(r) \left( \frac{X_0(r)}{X(t,r)} \right)^2 \right] \ .
\end{equation}

The main difference from the homogeneous Friedmann equation is that LTB generalization~\eqref{Hubble} has space and time dependence. All the LTB quantities depend not only on time but also on the radial coordinate $r$. Notwithstanding, the whole formalism is covariant under radial coordinate re-definition. Indeed, the LTB metric~\eqref{LTBmetric} and all the formulae are covariant under the change $r\rightarrow f(r)$. Therefore, by a suitable choice of radial coordinate one can choose freely the value of the scale function today $X_0(r)$. A convenient choice that shall be assumed in what follows is $X_0(r)=r$. This gauge fixation is similar to the normalization of the scale factor today for a FLRW universe, i.e. choosing $a(t_0)=1$.

The Friedmann-like equation \eqref{Hubble} can be integrated to give the age parameter $\Delta t(r)$. The spatial inhomogeneity of the metric induce a spatial dependent age parameter. It is convenient to fix the age parameter as the time spanned since the Big Bang time $t_B(r)$ which is defined as the time when $X(t_B,r)=0$. Thus, integrating \eqref{Hubble} gives
\begin{eqnarray}\label{t-tB}
\Delta t(r)= t-t_B(r)&=& \frac{1}{H_{\bot 0}(r)} \int ^{x(t,r)}_0 \frac{d {y}}{\sqrt{{y}^{-1} \Omega _M  + \Omega _K }} \ ,
\end{eqnarray}
where $x(t,r) =X(t,r)/X_0(r) =r^{-1}\, X(t,r)$. The Big Bang time works as a constant of integration and as such is a free function of the model. The model is formally specified once the matter density parameter $\Omega _M (r)$ and the Big Bang time $t_B(r)$ are given. In principle one could choose arbitrarily these two functions at each spatial location. However, it has been shown \cite{silk1977, zibin2008} that even a small spatial dependence on the Big Bang time can produce too large inhomogeneities today to agree with the observed CMB. Thus, it is commonly assumed a simultaneous Big Bang time such that $t_B(r)=t_\ast$ with $t_\ast$ a constant. With the hypothesis of a simultaneous Big Bang, the numerical value of $t_\ast$ is not important and we can set it to zero. Equation \eqref{t-tB} calculated today gives
\begin{eqnarray}
H_{\bot 0}(r) &=& \frac{1}{t_0} \int ^{1}_0 \frac{d {y}}{\sqrt{{y}^{-1} \Omega _M  + \Omega _K }} \ , \label{H0}
\end{eqnarray}
where $t_0$ is the age of the universe which is chosen to be $t_0 = 13.7$ Gyr\footnote{Comparing to the best fit $\Lambda$CDM model, that corresponds to fix $H_0 = 71 \mbox{ Km}/ \mbox{Mpc}/ \mbox{sec}$.}. Note that $H_{\bot 0}(r)$ is measured in units of $\mbox{Gyr}^{-1}$. For each $\Omega _M $, equation \eqref{H0} allows us to calculate the Hubble factor today $H_{\bot 0}(r)$ which then can be used in \eqref{Hubble} to generate the scale factor $X(t,r)$ and all its derivatives for each radius and time.

The above procedure yields the background dynamics. In order to compare the LTB model with observations, we also need to construct the light paths. This is accomplished by solving the appropriate null geodesics for the model. The LTB spacetime is spherically symmetric which implies that an observer at its center of symmetry ($r=0$) shall measure incoming radial trajectories. Thus, we can consider with full generality radial geodesics to follow a line with $\dd \theta = \dd\varphi = 0$. The null geodesics has vanishing interval, i.e. $\dd s^2=0$, hence, \eqref{LTBmetric} show us that for null geodesics we have 
\begin{eqnarray}
\frac{\dd t}{\dd \lambda} &=& -\frac{\dd r}{\dd \lambda} \frac{X^{\prime}(t,r)}{\sqrt{1+2E(r)}} \ , \label{geodnull}
\end{eqnarray}
where $\lambda$ is an affine parameter and we kept the minus sign inasmuch we are considering incoming trajectories. It can be shown~\cite{Enqvist} that for two successive light rays, obeying the above radial null geodesic and emitted respectively at time $t_1$ and $t_1+\delta t$, the period between wavefronts satisfies
\begin{eqnarray}
\frac{\dd \delta t}{\dd \lambda} &=& -\frac{\dd r}{\dd \lambda} \frac{\dot{X}^{\prime}(t,r)\delta t(\lambda)}{\sqrt{1+2E(r)}} \ . \label{geodnullperiod}
\end{eqnarray}

Recalling the redshift definition, namely $z(\lambda)=\nu_0/\nu(\lambda)-1$, \eqref{geodnullperiod} can be recast as
\begin{eqnarray}
\frac{\dd z}{\dd \lambda} &=& (1+z)\frac{\dd r}{\dd \lambda} \frac{\dot{X}^{\prime}(t,r)}{\sqrt{1+2E(r)}} \ . \label{tolrel}
\end{eqnarray}

Equations \eqref{geodnull} and \eqref{tolrel} determine the null geodesic equations in terms of the redshift
\begin{eqnarray}
\frac{dt}{dz} &=& -\frac{X^{\prime}(t,r)}{(1+z)\dot{X}^{\prime}(t,r)} \ ,\label{dtdz} \\ 
\frac{dr}{dz} &=& \frac{c\sqrt{1+2E(r)}}{(1+z)\dot{X}^{\prime}(t,r)} \ , \label{drdz}
\end{eqnarray} 
where we have explicitly re-introduced the speed of light  $c\approx 0.3$ Gpc/Gyr\footnote{Note that writing the speed of light in units of Gpc/Gyr we obtain the luminosity distance in Gpc since we have set our time scale in Gyr.}.

In order to solve \eqref{dtdz} and \eqref{drdz} we need two initial conditions. A suitable choice is the point at $z=0$ given by $\big(t(0)=t_0, r(0)=0\big)$. Thus, solving the system \eqref{dtdz} and \eqref{drdz} we have the light curve $\big(t(z), r(z)\big)$. 

In a LTB universe, the angular diameter distance measured by an observer at the center is directly related to the scale function. With  the procedure described above, we have the scale function at every point $X(t,r)$ and the radial null trajectory $\big(t(z), r(z)\big)$, hence, we can follow the scale function throughout the geodesics and obtain the angular diameter distance as a function of the redshift 
\begin{eqnarray}
d_A^\sube{LTB}(z) &=& X \left( t(z), r(z) \right) \ .\label{dA}
\end{eqnarray} 

In addition, we can also calculate the luminosity distance directly through its relation with the angular diameter distance~\cite{Etherington,Ellis}, namely
\begin{eqnarray}
d_L^\sube{LTB}(z) &=& (1+z)^2 d_A^\sube{LTB} (z) \ .\label{dL}
\end{eqnarray}

Finally, in order to relate the inhomogeneous LTB model with the SNIa observations we also need the distance modulus
\begin{eqnarray}
\mu ^\sube{LTB}_{th}(z) \equiv m - M = 5 \log _{10} \left( \frac{d_L^\sube{LTB}(z)}{10 \mbox{pc}} \right) \ ,
\end{eqnarray}
where $m$ is the apparent magnitude of a source with absolute magnitude $M$. For every matter density profile $\Omega_M(r)$ we can run the above scheme and fit the observational data.

\section{The void profiles}\label{sec:void_profiles}

As it is well known, in the $\Lambda$CDM model the observed dimming of distant supernovae can only be explained by the ad hoc hypothesis of a dark energy component responsible for driving the recent accelerating expansion of the universe. However, within spherically symmetric inhomogeneous models, it is possible to generate the observed dimming of distant objects via a local underdense region. 

\begin{figure}
\centering
\includegraphics[width=10cm,height=5cm]{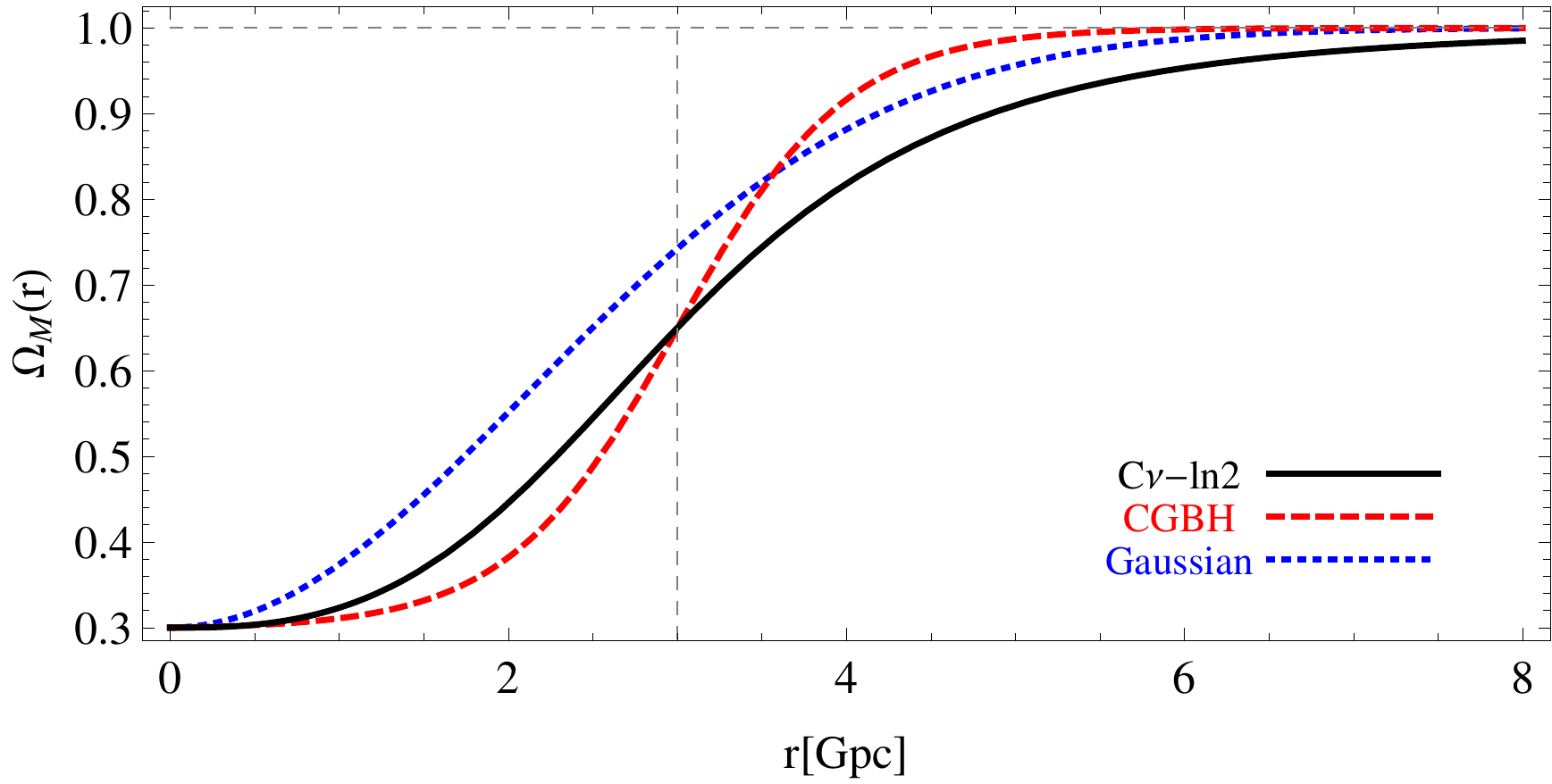} 
\caption{Matter density parameter today as a function of the radial coordinates. In all three profiles the outer density parameter is chosen unity $\Omega_{M,\scriptsize{\mbox{out}}}=1$. The others are free parameters to be adjust with observational data. To plot the profiles we used $\Omega_{M,\scriptsize{\mbox{in}}}=0.3$, $\Delta r = \nu = 0.5$ and $r_0=3$.}\label{figuravoid}
\end{figure}

In the present work, we analyze three different profiles of matter distributions, namely the CGBH, gaussian-like and the C$\nu$-ln2 profile which we describe below. All three profiles share two main properties. They describe a local spherically symmetric underdense vicinity that smoothly approach unity in the faraway region (see Fig.\ref{figuravoid}). The parameters of each profile is chosen such as to recover asymptotically the FLRW spacetime. This asymptotic behavior also guarantees that in the far past all models approach a FLRW universe.

In~\cite{GBH}, J.~Garcia-Bellido and T.~Haugb\o lle proposed a model with six parameters that is completely characterized by the matter density $\Omega_M(r)$ and the transverse expansion rate today $H_{\bot 0}(r)$. Their parametrization fixes the inner and outer values of $\Omega_M$ and $H_{\bot 0}$ and also how large and smooth is the transition from the inner and outer regions. They have also considered a more constrained profile (CGBH) by requiring a simultaneous Big Bang time. This extra condition impose a relation between the expansion rate and the matter density and hence, this model has only a single free function.

The CGBH profile can be parameterized as
\begin{eqnarray}\label{profile_CGBH}
\Omega _M(r) &=& \Omega _{M\sube{out}} + (\Omega _{M\sube{in}} - \Omega _{M \sube{out}} ) \left[  \frac{1- \tanh \left[ (r-r_0)/ 2 \Delta r \right]}{1+ \tanh(r_0/2\Delta r)} \right] \ ,
\end{eqnarray}
where $\Omega_{M\sube{in}}$ is the matter density at the center of the void and $r_0$ is the typical size of the void. The $\Delta r$ parameter controls the steepness of the transition from inside to outside the void. In order to well adjust the observational data, the typical size of the void in LTB models is of the order of Gpc. The last parameter $\Omega_{M\sube{out}}$, as previously mentioned, was fixed to unity in order to asymptotically recover the flat FLRW model. The second profile is similar to the CGBH but has one parameter less and displays a gaussian-like transition from inside the void to the outer region. The matter density for the Gaussian profile reads
\begin{eqnarray}
\Omega _M(r) &=& \Omega _{M\sube{out}} + (\Omega _{M\sube{in}} - \Omega _{M\sube{out}} ) e ^{- \left(\frac{r}{r_0} \right)^2}  \ .
\end{eqnarray}

All the parameters have the same physical interpretation as before.
\begin{figure}
\centering
\includegraphics[width=10cm,height=5cm]{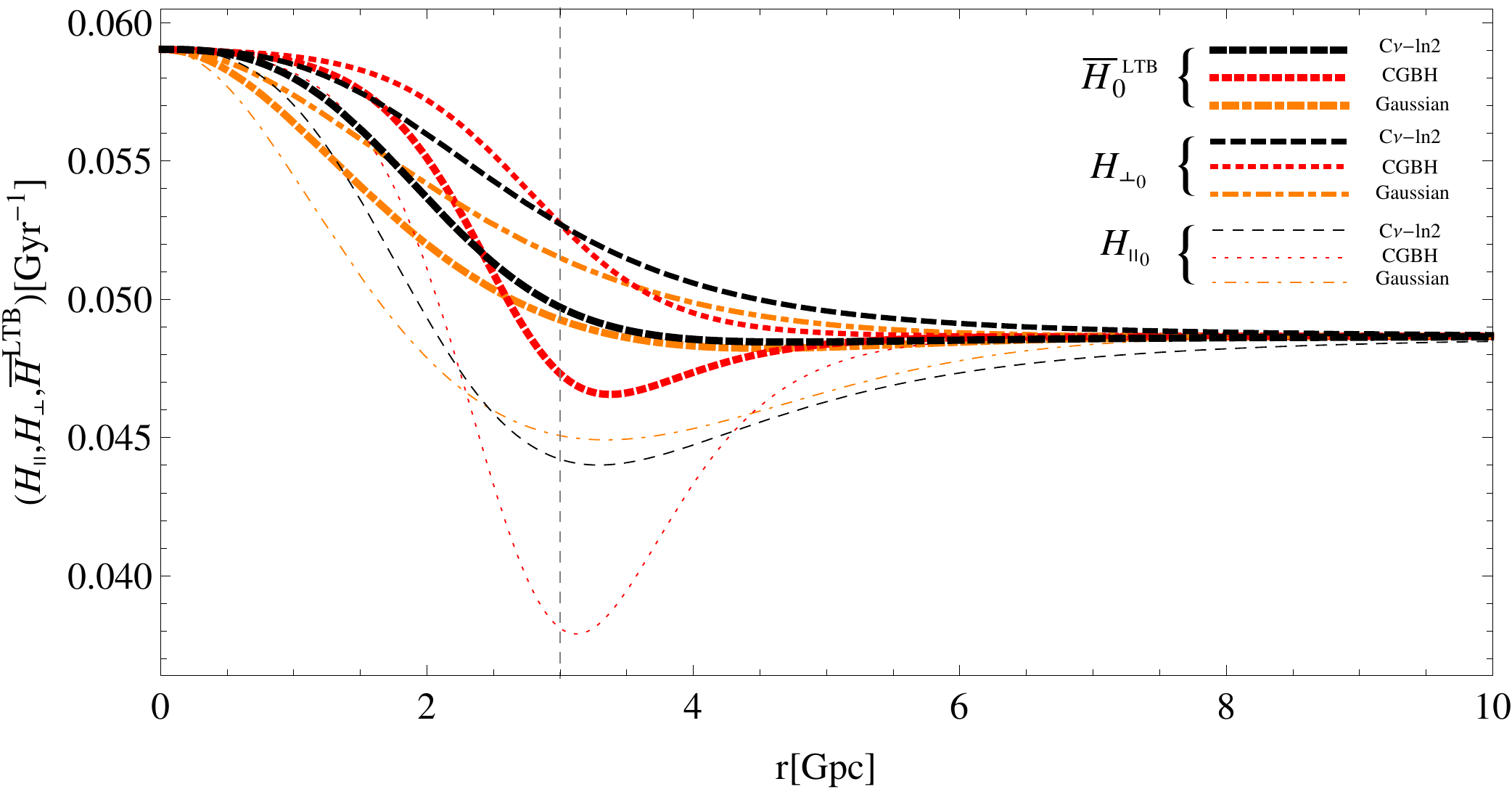}
\caption{
Plot of the three expansion rates today: transverse $H_{\bot 0}(r)$, the radial $H_{\parallel 0}(r)$ and the geometrical mean $\bar{H}_\sube{LTB}(t_0,r)$. The parameters used to plot the three profiles was $\Omega_{M\sube{out}}=1$, $\Delta r=\nu =0.5$, $\Omega_{M\sube{in}}=0.3$ and $r_0=3$.}\label{rates}
\end{figure}
The third and last profile follows a similar reasoning of the previous parametrization. It is a void model that smoothly changes the matter density from an underdense environment with $\Omega_{M\sube{in}}$ to an outer denser region with matter density $\Omega_{M\sube{out}}$. The profile reads
\begin{eqnarray}\label{profile_Cnuln2}
\Omega_M(r) &=& \Omega _{M\sube{in}} + (\Omega _{M\sube{out}}-\Omega _{M\sube{in}}) \frac{\left(\frac{r}{r_0} \right)^{3 + \nu} \Big[ 1 + \nu \ln ^2 \left(\frac{r}{r_0}\right) \Big]}{1+ \left(\frac{r}{r_0} \right)^{3+\nu} \Big[ 1 + \nu \ln ^2\left(\frac{r}{r_0}\right) \Big] }\ , 
\end{eqnarray}
where $\nu$ plays a similar role as $\Delta r$ for the CGBH profile. A nice feature that distinguishes the above profile from the CGBH is the independence of the value of the matter density at $r_0$ with respect to the transition parameter. In the C$\nu$-ln2 parametrization $ \Omega_M(r_0)$ depends only on $\Omega_{M\sube{in}}$ and $\Omega_{M\sube{out}}$. Indeed, a direct calculation shows that
\begin{eqnarray}
\Omega_M(r_0) &=& \frac{\Omega_{M\sube{in}}-\Omega_{M\sube{out}}\tanh\left(r_0/2\Delta r\right)}{1+ \tanh\left(r_0/2\Delta r\right)}\qquad \qquad \mbox{CGBH profile}\\
\Omega_M(r_0) &=& \frac12\left(\Omega_{M \sube{out}}+\Omega_{M \sube{in}}\right)\ \qquad \qquad \qquad \qquad \mbox{C$\nu$-ln2 profile}
\end{eqnarray}

In Fig.~\ref{rates}, we depict the transverse and radial expansion rates~\eqref{Hsparam} and the geometrical mean $\bar{H}^\sube{LTB}$ for these three different profiles. It is worth noting that for the CGBH model the difference between the transverse and radial expansion rates is prominently larger than in the other models.

\subsection{Deceleration parameter}\label{subsec:decelparam}

The two expansion rates in a LTB model, viz. the transverse $H_{\bot 0}(r)$ and the radial $H_{\parallel 0}(r)$ expansion rates, can be used to define two different deceleration parameters. Thus, in analogy to the FLRW metric, we can define the transverse and radial deceleration parameter as 
\begin{eqnarray}
q_{\bot }(t,r) &=& - \frac{\ddot{X}(t,r)}{X(t,r)H^{2}_{\bot }(t,r)} \quad  , \\
q_{\parallel }(t,r) &=& - \frac{\ddot{X}^{\prime}(t,r)}{X^{\prime}(t,r)H^{2}_{\parallel}(t,r)}   \quad  .
\end{eqnarray}  

\begin{figure}
\centering
\includegraphics[width=8cm,height=4.5cm]{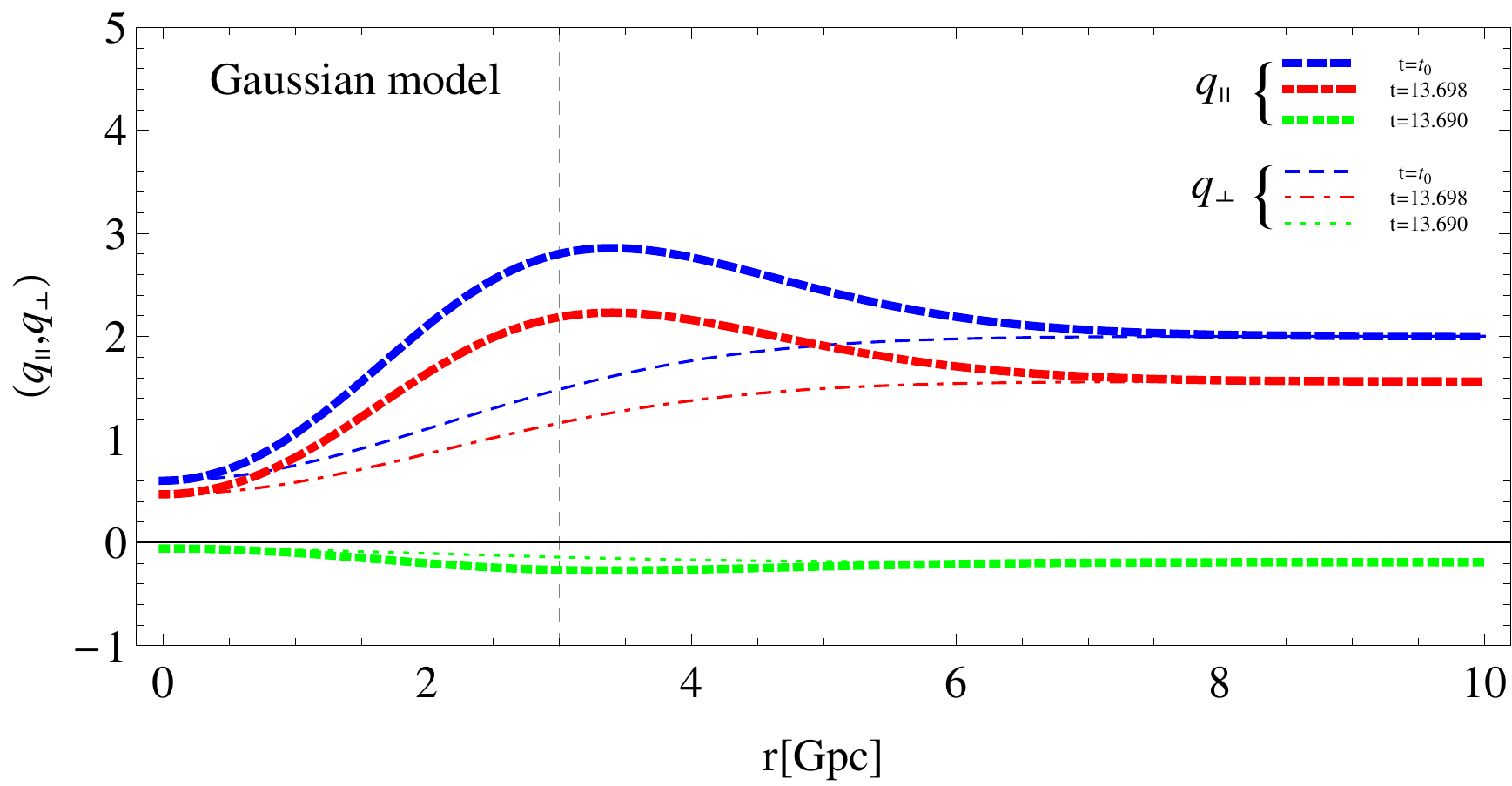}
\includegraphics[width=7.5cm,height=4.5cm]{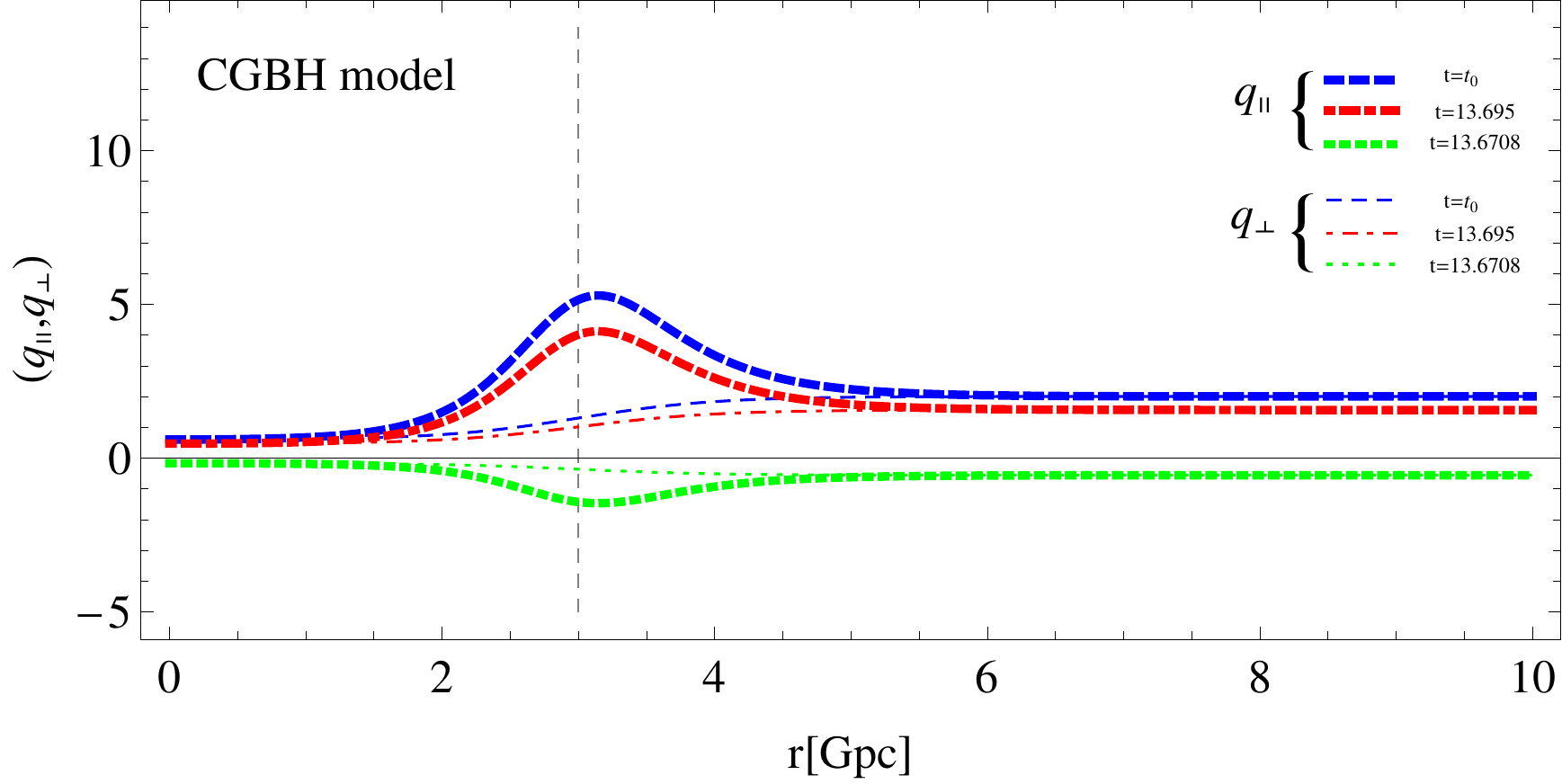}
\includegraphics[width=7.5cm,height=4.5cm]{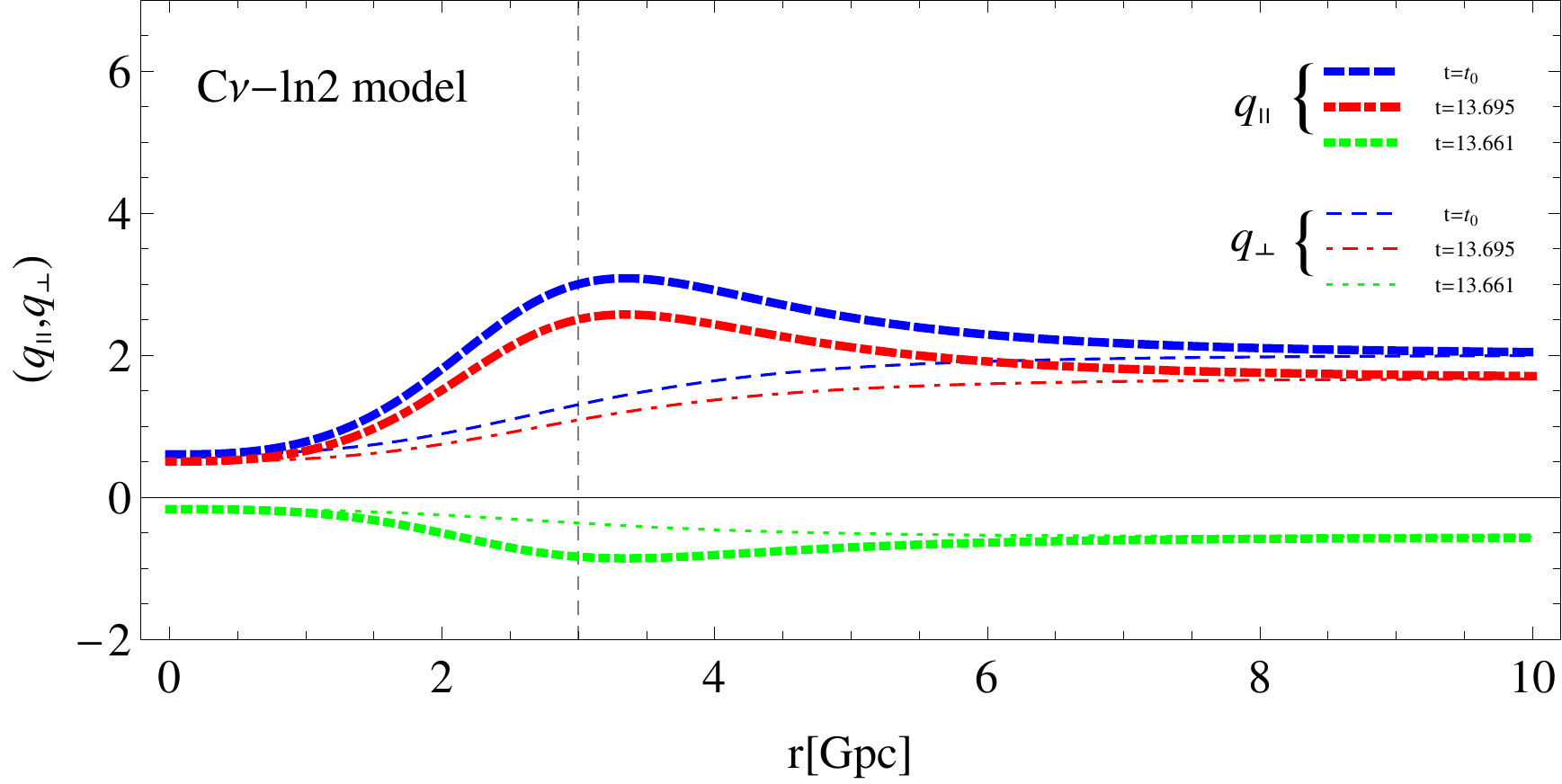}
\caption{Plot of the radial and transverse  deceleration parameters as a function of $r$ at different times. The vertical dashed-line mark the typical size of the void $r_0$. Note that close to time $t_0$ all the curves are positive, hence, displaying deceleration universes today.} \label{decelrt0}
\end{figure}

The LTB solution reduces to a FLRW metric when $X(r,t) = a(t)\, r $, with $a(t)$ being the scale factor. One can immediately verify that in this case both deceleration parameter defined above merge into the FLRW parameter, i.e. $q=q_{\parallel} = q_{\bot} = -\frac{\ddot{a}}{a H^2}$. Figure~\ref{decelrt0} shows both deceleration parameters as a function of radius at different times. 

In all cases the expansion today is decelerated contrasting with the $\Lambda$CDM model that describes a late time accelerating universe. This is a general feature of LTB models which is considered in the literature as a powerful manner to distinguish a homogeneous and isotropic evolution from inhomogeneous models~\cite{timedrift, redshiftdrift}.

Besides the two deceleration parameter defined above, we can still combine them to form an effective deceleration parameter. In LTB spacetimes the expansion factor of a time-like congruence of static observers with $v^\mu=\delta^\mu_0$ is given by $\Theta =\nabla_\mu v^\mu=H_{\parallel}+2H_{\bot}$. On the other hand, in a FLRW universe we can write the deceleration parameter as $q=-1-3\dot{\Theta}/\Theta$. Therefore, we define an effective LTB deceleration parameter simply by replacing the expansion factor for its LTB version. Using \eqref{dtdz}, the effective deceleration parameter in terms of the redshift reads
\begin{eqnarray}
q^\sube{eff}(z) &=& -1 +  \frac{3(1+z)H_{\parallel}(z)}{[H_{\parallel}(z) + 2 H_{\bot}(z)]^2} \left[ \frac{d}{dz}H_{\parallel}(z) + 2 \frac{d}{dz}H_{\bot}(z)\right] \ .
\end{eqnarray}

\begin{figure}
\centering
\includegraphics[width=10cm,height=5cm]{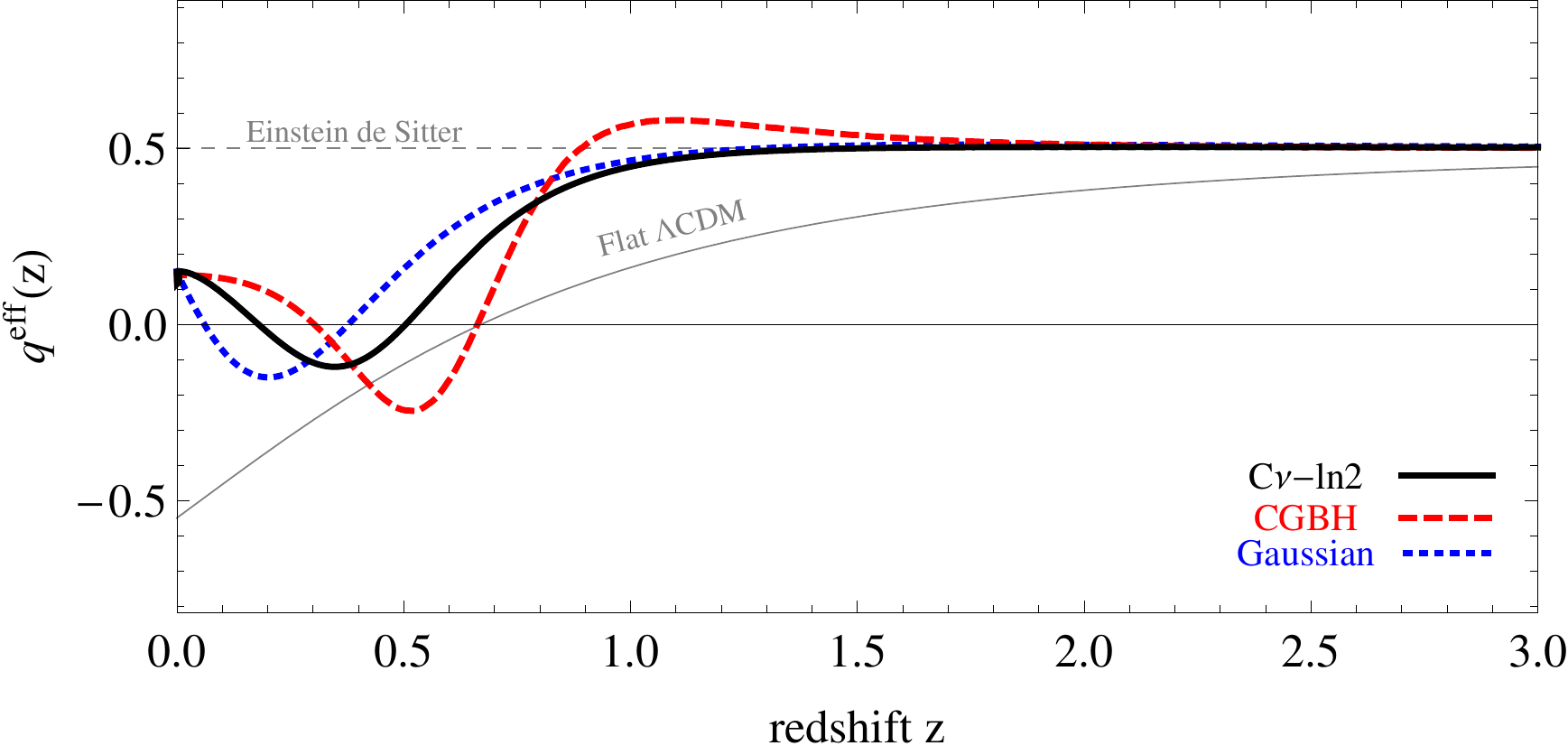}
\caption{Effective deceleration parameter for the 3 models compared with the flat $\Lambda$CDM and the EdS model. } \label{figqeff}
\end{figure}

Note that for large redshift the effective parameter approaches the FLRW behavior. Figure~\ref{figqeff} shows the evolution in redshift of the effective deceleration parameter for the three profiles. Additionally, we included for comparison the flat $\Lambda$CDM (with $\Omega_m = 0.3 $) and the Einstein-de Sitter (EdS) models. The deceleration parameter for all inhomogeneous models changes sign twice displaying decelerating expansion for small $z$. This behavior is in agreement with the transverse and parallel deceleration parameters describe above.

\subsection{BAO in LTB models}\label{subsec:BAOinLTB}

The standard model describes the primordial nucleosynthesis and recombination processes within a homogeneous and isotropic universe. The BAO is a characteristic length scale imprinted in the matter distribution that encodes the physics of this early phase of the universe. This small signal excess occurs at length separations of the order of 150 Mpc.

In order to deal with BAO observations in inhomogeneous models, one has to tackle different effects coming from the spatial dependence of the background dynamics. In particular, in a LTB model the physical length depends not only on time but also on the radial coordinates. Furthermore, the model has two distinct expansion rates, i.e. the transverse $H_{\bot}(t,r)$ and the radial $H_{\parallel}(t,r)$ expansion rates. In an arbitrary inhomogeneous model, one has to start the BAO analysis from first principles. However, the profiles \eqref{profile_CGBH}-\eqref{profile_Cnuln2} together with the hypothesis of a simultaneous Big Bang time guarantee that our LTB models have the valuable property of approaching the FLRW dynamics in the far past. Therefore, we can assume that these models are indistinguishable from a FLRW universe during recombination. The crucial step is then to properly propagate the initial condition through the inhomogeneous LTB dynamics. The transverse and radial components evolve differently, hence, following~\cite{zumalacarregui}, we should describe them separately.

The spherical symmetry of the LTB metric considerably simplifies the evolution of the transverse length. Indeed, one can choose the coordinate systems such that $\dot{\theta}=\dot{\varphi}=0$. In addition, these conditions are stable for timelike geodesics. Therefore, an initial angular separation $\dd \theta$ between neighbor geodesics is preserved along the trajectory. Thus, the angular physical length at emission is given by $L_{\bot}(t_e,r) =  \int \sqrt{g_{\theta \theta}}\, \dd\theta = X(t_e,r)\, \dd\theta$. The radial physical length is obtained in a similar manner. A material particle initially at rest will continue at the same radial position and this evolution is again stable for timelike geodesics. For an initial radial separation $\dd r$, the radial physical length at emission is $L_{\parallel}(t_e,r) = \int \sqrt{g_{r r}}\, \dd r \approx X^{\prime}(t_e,r)\, \dd r/\sqrt{1+2E(r)}$. These lengths can be related to the observed physical lengths respectively as 
\begin{eqnarray}
L_{\bot}(t,r) &=& \frac{X(t,r)}{X(t_e,r)}L_{\bot}(t_e,r) \ ,\label{LphysTrans} \\
L_{\parallel}(t,r) &=& \frac{X^{\prime}(t,r)}{X^{\prime}(t_e,r)} L_{\parallel}(t_e,r) \ .\label{LphysRad}
\end{eqnarray}

The above relations express the background evolution of the physical length in LTB model. The first order corrections are not so straightforward as in the homogeneous and isotropic case. In a FLRW universe, the evolution of sub-horizon first order perturbations is exclusively time dependent, hence, spatial dependence on BAO scale comes only from non-linear corrections. Nowadays cosmological observations precision is accurate enough to urge for an adequate control of these effects. As argued in \cite{crocce2008,angulo2008,smith2008} these effects can produce a shift in the acoustic scale which in turn might generate systematics compared to expected statistical errors in the next generation of surveys. Notwithstanding, for the present analysis we shall keep only up to first order perturbations.

The evolution of linear perturbations in inhomogeneous models is more complicate than in FLRW. One of the key issues is the mixing of first order perturbation modes. As it is well known
\cite{zibin2008,clarkson2009}, the decoupling of linear order perturbations is a direct consequence of the symmetries of the background FLRW metric. In a LTB universe it is expected a coupling between the scalar, vector and tensor modes. The crucial coupling in a LTB universe is between the scalar and tensor modes. Indeed, due to the spherical symmetry of the background LTB metric, it can be shown that vector modes can be disregarded compared to the other components. Furthermore, the mode-coupling depends on the specific matter profile of the model. Fortunately, for the CGBH profile these nonlinear effects are subdominant~\cite{zumalacarregui,february2012} and one can approximately consider the BAO scale to be constant in coordinate space. The validity for the gaussian and C$\nu$-ln2 profiles follows from their lower difference between the radial and transverse expansion rates (see Fig.~\ref{rates}). 

The LTB models considered here approach the dust FLRW behavior either for large spatial distances ($r=r_\infty\gg r_0$) or far in the past ($t=t(z_\infty)$ with $z_\infty \gtrsim 1000$). Therefore, we can assume that early times baryonic dynamics in our model is indistinguishable from the FLRW case. This feature guarantees the use of the spatial independent Big Bang BAO scale as initial condition for our models. Additionally, in the far past, the approximate homogeneity and isotropy of the universe allows us to assume the BAO scale to be spatial coordinate independent at constant time hypersurfaces 
\begin{eqnarray}
L^\sube{BAO}\big(t_e,r(z)\big) & \approx &  L^\sube{BAO}(t_e,r_{\infty}) \ . \label{LBAOte}
\end{eqnarray}

The asymptotic BAO scale can be computed within a $\Lambda$CDM scenario for which we use the fitting formulae for an adiabatic cold dark matter developed in~\cite{Eisenstein}. The asymptotic BAO scale depends on the sound horizon $l_{s}$ at the drag epoch which can be approximated by
\begin{eqnarray}\label{horzsound}
l_{s}(z_\sube{drag})&=& \frac{44.5  \ln(9.83/\Omega^\sube{eff}_{M} h^{2}_\sube{eff})}{\sqrt{1+10(\Omega^\sube{eff}_b h^2_\sube{eff})^{3/4}}} \ \mbox{Mpc} \ ,
\end{eqnarray}
where $\Omega^\sube{eff}_{M}$, $\Omega^\sube{eff}_b$ and $h_\sube{eff}$ are respectively the effective total density, effective baryon density and the effective normalized Hubble factor. In a conventional $\Lambda$CDM model we would have instead the total matter parameter $\Omega_0$, the baryonic parameter $\Omega_b$ and the normalized Hubble factor $h=H_0/(100 \mbox{ Km} . \mbox{Mpc}^{-1}. \mbox{s}^{-1})$. However, since in the asymptotic regime our model is only approximately homogeneous and isotropic we need to evolve backwards in time the LTB dynamics up to the time of emission and use the effective values in the above fitting formulae. 

The emission time has to be far enough in the past to reach the asymptotic FLRW behavior. Due to computational economy we select the emission time as $z_e = 100$ when the LTB inhomogeneities\footnote{The inhomogeneities can be characterized through the density contrast $\delta\rho_m(t,r)= \rho_m(t,r)/\rho_m\big(t_\infty,r\big)$, hence, the statement that LTB inhomogeneous are small means that $\delta \rho(t,r)\lesssim 1\%$.} are still of the order of $1\%$. At this stage we can assume $\Omega^\sube{eff}_{M} = \Omega_{M\sube{out}}$ and $\Omega^\sube{eff}_b = f_b\,  \Omega_{M\sube{out}}$ where $f_b$ is the fraction of baryon to total matter. The normalized Hubble factor reads $h_\sube{eff}=H^\sube{eff}_0/(100 \mbox{ Km} . \mbox{Mpc}^{-1}. \mbox{s}^{-1})$ where the effective Hubble parameter is given by
\begin{eqnarray}
H^\sube{eff}_0 = \frac{\left[  H_{\parallel}(z_e)H^2_{\bot}(z_e) \right]^{1/3}}{\sqrt{\Omega^\sube{eff}_{M}(1+z_e)^3+(1-\Omega^\sube{eff}_{M})(1+z_e)^2}} \ .
\end{eqnarray}

The single difference in the above procedure from~\cite{zumalacarregui} is the use of the geometrical mean Hubble parameter \eqref{HLTBmean}, instead of the arithmetic mean Hubble parameter \eqref{HLTBmean_arith}.
This modification has the advantage of the geometrical mean converge faster to the FLRW  regime than the arithmetic mean.

The sound horizon gives the comoving BAO scale which in a flat FLRW universe equals the physical BAO scale today\footnote{The identification of the comoving sound horizon with the physical BAO scale today is valid if we assume the normalization of the scale factor $a(t_0)=1$.}. Recalling that the asymptotic spatial limit of the LTB model at any time is a FLRW universe, we have $L^\sube{BAO}(t_0,r_{\infty})=l_{s}(z_\sube{drag})$. In order to relate the sound horizon~\eqref{horzsound} with the radial and transverse physical scales using~\eqref{LphysTrans}-\eqref{LBAOte} one perform a cross-multiplication to obtain
\begin{eqnarray}
L^\sube{BAO}_{\parallel}(z) &=& \frac{X^{\prime}\big(t \mbox{{\small (}}z\mbox{{\small)}},r\mbox{{\small (}} z \mbox{{\small)}}\big)}{X^{\prime}\big(t_e,r\mbox{{\small(}}z\mbox{{\small)}}\big)} \frac{X^{\prime}\big(t_e,r_{\infty}\big)}{X^{\prime}\big(t_0,r_{\infty}\big)} l_{s}(z_\sube{drag}) \ , \label{LBAOparl} \\
L^\sube{BAO}_{\bot}(z) &=& \frac{X\big(t \mbox{{\small (}}z\mbox{{\small)}},r\mbox{{\small (}} z \mbox{{\small)}}\big)}{X\big(t_e,r\mbox{{\small(}}z\mbox{{\small)}}\big)} \frac{X\big(t_e,r_{\infty}\big)}{X\big(t_0,r_{\infty}\big)} l_{s}(z_\sube{drag}) \ . \label{LBAOtrans}
\end{eqnarray}

There is one last step to connect the above relations to the observational data. The sensitivity of current surveys provide only a combined distance scale ratio from the spherically averaged power spectrum~\cite{okumura2008,shoji2009}. In a FLRW universe, the physical observable associated with the BAO scale is the ratio
\begin{equation}\label{baoobsFLRW}
\theta_\sube{FLRW}=\frac{l_{s}(z_\sube{drag})}{D^\sube{FLRW}_V(z)}\ ,
\end{equation}
where $D^\sube{FLRW}_V(z)$ encodes the dilation scale as the cube root of the product of the radial dilation with the square of the transverse dilation~\cite{Eisenstein2005,Percival2010}. The radial dilation is given by $D^\sube{FLRW}_z(z)=z/H^\sube{FLRW}(z)$ whereas the angular or transverse dilation is simply the comoving angular diameter distance that can be written in term of the diameter distance as $(1+z)D^\sube{FLRW}_A$. Thus, we write 
\begin{equation}
D^\sube{FLRW}_V(z)=\left[\Big((1+z)D^{\sube{FLRW}}_A\Big)^2\frac{z}{H^\sube{FLRW}(z)}\right]^{1/3}\ .
\end{equation}

Even though \eqref{baoobsFLRW} gives a BAO observable, actual measurements usually refer to the model-independent quantities $(\Delta \theta^2\Delta z)$ where $\Delta \theta$ is the angular in the sky and $\Delta z$ is the redshift interval corresponding to the comoving sound horizon. In a FLRW universe we have $D_A=l_s/(1+z_\sube{BAO})\Delta \theta$ and $l_s=\Delta z/H(z_\sube{BAO})$ hence we find
\begin{equation}
(\Delta \theta^2\Delta z)^{1/3}=z^{1/3}\, \frac{l_s}{D_V(z)}\qquad \mbox{for FLRW}\ .
\end{equation}

In accordance with~\cite{zumalacarregui,Biswas2010}, we shall define in the LTB model a characteristic BAO length $d_z(z)$ as the LTB analogue of the above relation. Thus, we define the length 
\begin{eqnarray} \label{dzBAOcorrel}
d_z (z)&=& \left( \frac{\Delta \theta ^2 \Delta z}{z} \right)^{1/3} \qquad \mbox{for LTB} \ .
\end{eqnarray}
The angular scale is again related to the angular diameter distance with the difference that we must use the transverse BAO length 
\begin{eqnarray}
\Delta \theta _\sube{BAO} &=& \frac{L^\sube{BAO}_{\bot}(z)}{d^\sube{LTB}_{A}(z)}  \label{DthetaBAO}.
\end{eqnarray}
In a similar manner, from~\eqref{drdz} and the metric~\eqref{LTBmetric}, the redshift separation reads
\begin{eqnarray}
\Delta z_\sube{BAO} &=& (1+z) H_{\parallel}(z)  L^\sube{BAO}_{\parallel}(z) \ . \label{DzBAO}
\end{eqnarray}
Thus, combining the above equations with~\eqref{LBAOparl} and~\eqref{LBAOtrans} we obtain
\begin{eqnarray}
d^\sube{LTB}_{z} &=& \left[\frac{1+z}{z} \frac{H_{\parallel}(z)}{d^\sube{LTB}_{A}(z)^{2}}  \right]^{1/3} \xi (z) \ l_{s}(z_\sube{drag}) \ , \label{dLTBz}
\end{eqnarray} 
where the redshift dependent function $\xi(z)$ reads
\begin{eqnarray}
\xi(z) &=& 
\left( \frac{X^{\prime}\big(t \mbox{{\small (}}z\mbox{{\small)}},r\mbox{{\small (}} z \mbox{{\small)}}\big)}{X^{\prime}\big(t_e,r\mbox{{\small(}}z\mbox{{\small)}}\big)} \frac{X^{\prime}\big(t_e,r_{\infty}\big)}{X^{\prime}\big(t_0,r_{\infty}\big)}\right)^{1/3} 
\left( \frac{X\big(t \mbox{{\small (}}z\mbox{{\small)}},r\mbox{{\small (}} z \mbox{{\small)}}\big)}{X \big(t_e,r\mbox{{\small(}}z\mbox{{\small)}}\big)} \frac{X\big(t_e,r_{\infty}\big)}{X\big(t_0,r_{\infty}\big)}\right)^{2/3} 
 \ .
\end{eqnarray}

\section{Observational data }\label{sec:obsdata}

\subsection{Type Ia supernovae}\label{subsec:snia_data}

In this paper we use the Joint Light-curve Analysis sample~\cite{Betoule} which is known in the literature as the JLA sample. This extended sample of 740 spectroscopically confirmed type Ia supernovae with high quality light curves consist of several low-redshift samples ($z<0.1$), the third year sample from the Sloan Digital Sky Survey (SDSS-II, $0.05<z<0.4$), the third year SuperNova Legacy Survey (SNLS, $0.2<z<1$) and the Hubble Space Telescope sample (HST, $z>1$). The observational distance modulus is modeled, in the context of the light curve fitter Spectral Adaptive Light curve Template (SALT2) \citep{guy07}, by
\begin{eqnarray}
\mu^\sube{SNIa}_{i} &=&  m^{\star}_{B,i}  + \alpha X_{1,i} - \beta C_{i}  - M_{B}
\end{eqnarray}
where $\alpha$, $\beta$ and $M_B$ are nuisance parameters in the distance estimate which are fitted simultaneously with the cosmological parameters. The absolute $B$-band magnitude is related to the host stellar mass ($M_\sube{stellar}$) by a simple step function
\begin{eqnarray}
M_{B} = \left\{
\begin{array}{lcl}
M^{1}_{B} \hspace*{1.2cm} && \quad \mbox{if} \ M_\sube{stellar} < 10^{10} \ M_{\odot} \  \\
&&\\
M^{1}_{B} + \Delta _{M}  &  & \quad  \mbox{otherwise} \ 
\end{array}
\right.
\end{eqnarray}
The light-curve parameters $\big(m^{\star}_B,\,  X_1,\,  C\big)$ result from the fitting of a model of SNe Ia spectral sequence to the photometric data. We can build the $\chi ^2$ function as
\begin{eqnarray}\label{chi2}
\chi ^2 (\boldsymbol{\theta},\boldsymbol{\delta},M_B) &=& \sum ^{740}_{i=1} \frac{\left[ \mu^\sube{SNIa}_{i}(\boldsymbol{\delta},M_B)-\mu ^\sube{LTB}_\sube{th}{}(z_{i};\boldsymbol{\theta}) \right]^2 }{\sigma ^{2}_{i} + \sigma ^{2}_\sube{int}}
\end{eqnarray}
where the supernovae parameters are denoted by $\boldsymbol{\delta} := (\alpha,\beta)$ and the cosmological parameters by $\boldsymbol{\theta}:=(\Omega_{M \sube{in}}, \Delta r ,\nu , r_0)$. The propagated error from the covariance matrix of the light-curve fitting is
\begin{eqnarray}
\sigma ^{2}_i &=& \sigma ^2_{m^{\star}_B,i} + \alpha ^{2} \sigma ^{2}_{X_1,i} + \beta ^{2} \sigma ^{2}_{C,i} + 2 \alpha \sigma _{m^{\star}_BX_{1},i} - 2 \beta \sigma _{m^{\star}_BC,i} - 2 \alpha \beta \sigma _{X_{1}C,i} + \sigma ^{2} _{\mu z , i} \ , 
\end{eqnarray}
where $\sigma ^{2}_{\mu  z,i}$ represents the contribution to the distance modulus coming from redshift uncertainties and peculiar velocities. Following~\cite{Ribamar,kessler2009} we shall simulate these effects by the distance-redshift relation for an empty universe, hence
\begin{eqnarray}
\sigma_{\mu z,i} &=& \sigma_{z ,i} \left(\frac{5}{\log 10}  \right) \frac{1+z_i}{z_i (1+ z_i/2) }  \ .
\end{eqnarray}

In the above expression $\sigma^2_{z,i} = \sigma^2_{\sube{spec},i} + \sigma^2_\sube{pec}$ with $\sigma_{\sube{spec} ,i}$ representing the redshift measurement error and $\sigma_\sube{pec}=0.0012$  is the uncertainty due to peculiar velocity. Finally, a floating term $\sigma_\sube{int}$ is also included in \eqref{chi2} to account for both intrinsic variations in the supernova luminosity and systematic effects.

Originally, the term $\sigma_\sube{int}$ was not considered as a free parameter to be optimized in the $\chi^2$ approach but should rather be determined by an iterative procedure. We start with a guess value for $\sigma_\sube{int}$ (usually around $0.15$) and perform the minimization procedure to obtain the best-fit values for the supernova and cosmological parameters. Then, with these best-fit values, we fine-tune $\sigma_\sube{int}$ such that the reduced $\chi^2$ goes to unity. The procedure is repeated with this new value of $\sigma_\sube{int}$ as input. The iteration ends when the value of $\sigma_\sube{int}$ converges.

The authors of~\cite{Ribamar} have exposed the limits of validity of the usual uncorrected $\chi^2$ approach. This happens when the covariance depends on the free parameters of the underlying model. In this case, one should use a parameter fitting based on the likelihood function 
\begin{eqnarray}
\mathcal{L}(\boldsymbol{\theta},\boldsymbol{\delta},M_B,\sigma_\sube{int}) &:=& \chi^2(\boldsymbol{\theta},\boldsymbol{\delta},M_B,\sigma_\sube{int}) + \sum ^N_i \ln(\sigma^2_i (\boldsymbol{\delta}) + \sigma^2_\sube{int}) \ ,  
\end{eqnarray}
where now $\sigma_\sube{int}$ is also considered as a free parameter. In this paper we shall follow both methods and compare their results in sec.~\ref{sec:results}.

\subsection{The BAO sample}\label{subsec:bao_sample}

The characteristic BAO scale detected in the correlation function of different matter distribution gives a powerful standard ruler to probe the angular diameter distance-redshift relation and the Hubble parameter. The BAO scale has been measured at different redshift values, namely at $z = 0.106$ by the 6dFGS~\cite{6dF}, $z = 0.35$ and $0.57$ by the SDSS~\cite{SDSS1,SDSS2}, $z = 0.44$, $0.6$ and $0.73$ by the WiggleZ collaboration~\cite{WiggleZ}. There is also an additional point presented by Carnero~et~al.~\cite{Carnero} at $z = 0.55$ indicating the angular correlation. All data are summarized in the Table~\ref{baodata}.
\begin{table}[H]
\centering
\begin{tabular}{|l|c|c|}
\hline \hline
sample & $z$ & $d_z \pm \sigma_{d_z}$ \\
\hline
6dFGS & 0.106 & $0.336 \pm 0.015$ \\
\hline
SDSS & 0.35 & $0.1126 \pm 0.0022$ \\
SDSS & 0.57 & $0.0732\pm 0.0012 $ \\
\hline
WiggleZ & 0.44 & $0.0916 \pm 0.0071 $ \\
WiggleZ & 0.6 & $0.0726 \pm 0.0034 $ \\
WiggleZ & 0.73 & $0.0592 \pm 0.0032 $ \\
\hline \hline
  sample        &  $z$  & $\Delta \theta \pm \sigma_{\theta}$ \\
\hline  
Carnero et al. & 0.55  &  $3.90^{o} \pm 0.38^{o}$ \\
\hline \hline
\end{tabular} 
\caption{BAO data summarized.}\label{baodata}
\end{table}

Then, the Likelihood approach is thus given by
\begin{eqnarray}
\chi ^2_{\tiny{\mbox{BAO}}} = \sum _{i,j} \left[d_{z,i} - d^{LTB}_{z}(z_i;\boldsymbol{\gamma}) \right] C^{-1}_{ij} \left[ d_{z,j} - d^{LTB}_z(z_j;\boldsymbol{\gamma}) \right] + \frac{\left[\Delta \theta - \Delta \theta _{BAO}(0.55)\right]^{2}}{\sigma ^{2}_{\theta}} \label{chi2Bao}
\end{eqnarray}
where $\boldsymbol{\gamma} := (\Omega_{M \sube{in}}, \Delta r ,\nu , r_0, f_b)$   and  $C^{-1}_{ij}$ is the inverse covariance matrix expressed in terms of the $d_z$:
\begin{eqnarray}
C^{-1}_{ij} =  
 \begin{pmatrix}
  4444 & 0. & 0. & 0. & 0. & 0. \\
0. & 206612 & 0 & 0. & 0. & 0. \\
0. & 0 & 694444  & 0. & 0. & 0. \\
0. & 0. & 0. & 23857 & -22747 & 10586 \\
0. & 0. & 0. & -22747 & 128729 &  -59907 \\
0. & 0. & 0. & 10586 & -59907 & 125536
 \end{pmatrix} \ .
\end{eqnarray}

\section{Results}\label{sec:results}

In this section we describe the result of the best fit parameters of the LTB model for the three profiles. Both the CGBH and C$\nu$-ln2 profiles have three parameters to specify its matter distribution plus an extra parameter $f_b$ in order to include the BAO analysis. In contrast, the Gaussian profile has only two parameters plus the $f_b$. Apart from this, there is also the supernovae parameters ($\alpha$, $\beta$, $M^1_B$ and $\Delta _M$) which, naturally, are the same for all cases. The numerical results are summarized in Tables~\ref{table2},~\ref{table3} and~\ref{table4}. For comparison, the table~\ref{tableLCDM} show the results of a corresponding analysis for the flat $\Lambda$CDM model setting $H_0 = 71$ Km/Mpc/sec.


Table~\ref{table2} shows the best fit parameters for the CGBH model using the $\chi ^2$ and Likelihood approaches. We display the best fit values separately for the JLA, the BAO and the JLA+BAO combined analysis. Table~\ref{table3} and~\ref{table4} show respectively the same information for the Gaussian and C$\nu$-ln2 profiles. Note that the supernovae parameters ($\alpha$, $\beta$, $M^1_B$ and $\Delta _M$) have similar results for the three cases. This is consistent with the idea that these parameters are not very sensitive to cosmological evolution. In contrast, the cosmological parameters ($\Omega_{\scriptsize{M,\mbox{in}}}$, $\Delta r$, $\nu$ and $r_0$) do show some appreciable differences. In particular, the C$\nu$-ln2  model requires the highest matter density within the void $\Omega_{\scriptsize{M,\mbox{in}}}$ relative to the CGBH and Gaussian models. This feature is robust in the sense that appears in both $\chi^2$ and Likelihood approaches.  

\begin{figure}
\includegraphics[width=9.cm,height=6.cm]{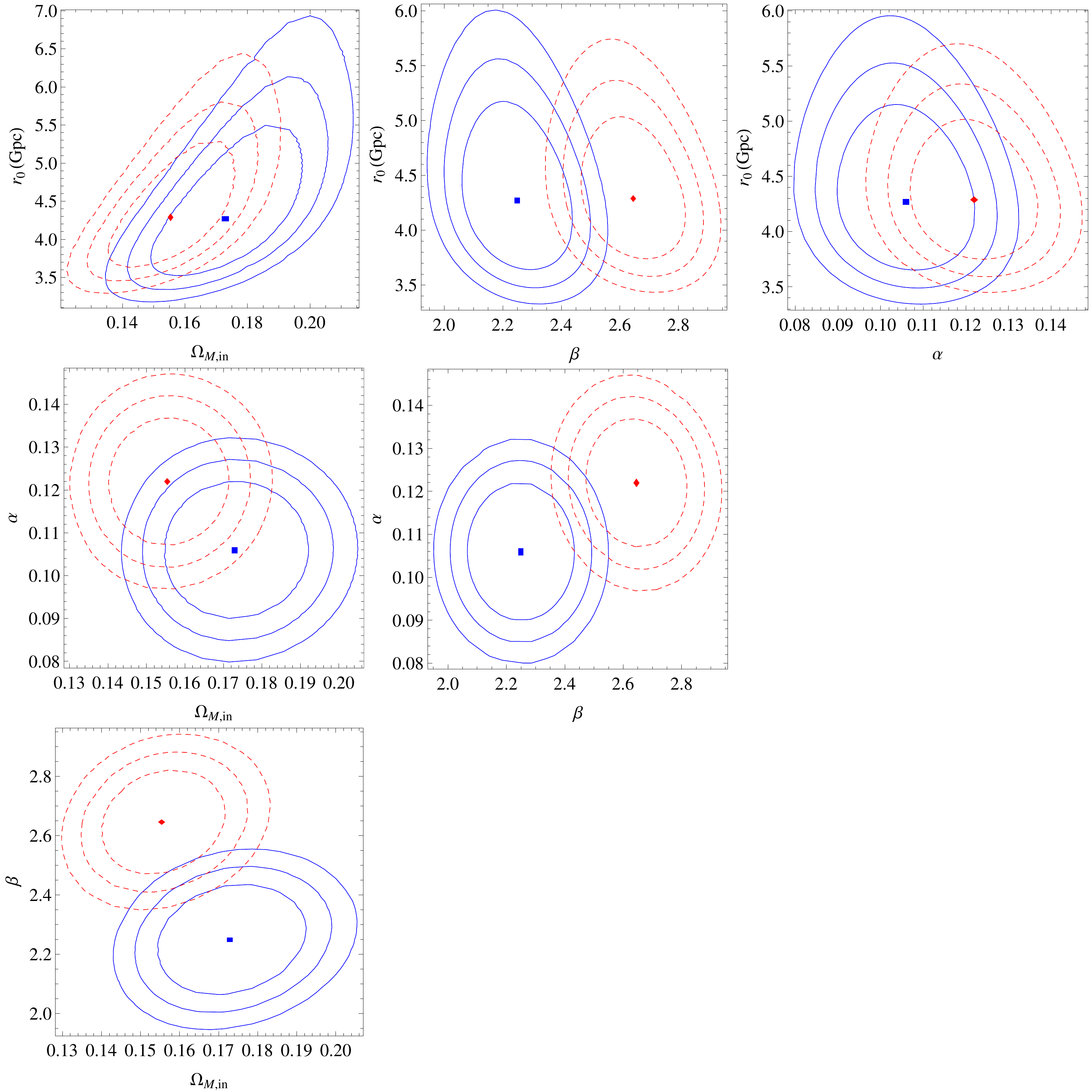}
\caption{Contour Level best fit parameters in the $\chi^2$ (dashed red) and Likelihood  (solid blue) approaches with the JLA sample only for the Gaussian model.  }  \label{fig:gaussian_sn}
\end{figure}

\begin{figure}
\includegraphics[width=12.cm,height=9cm]{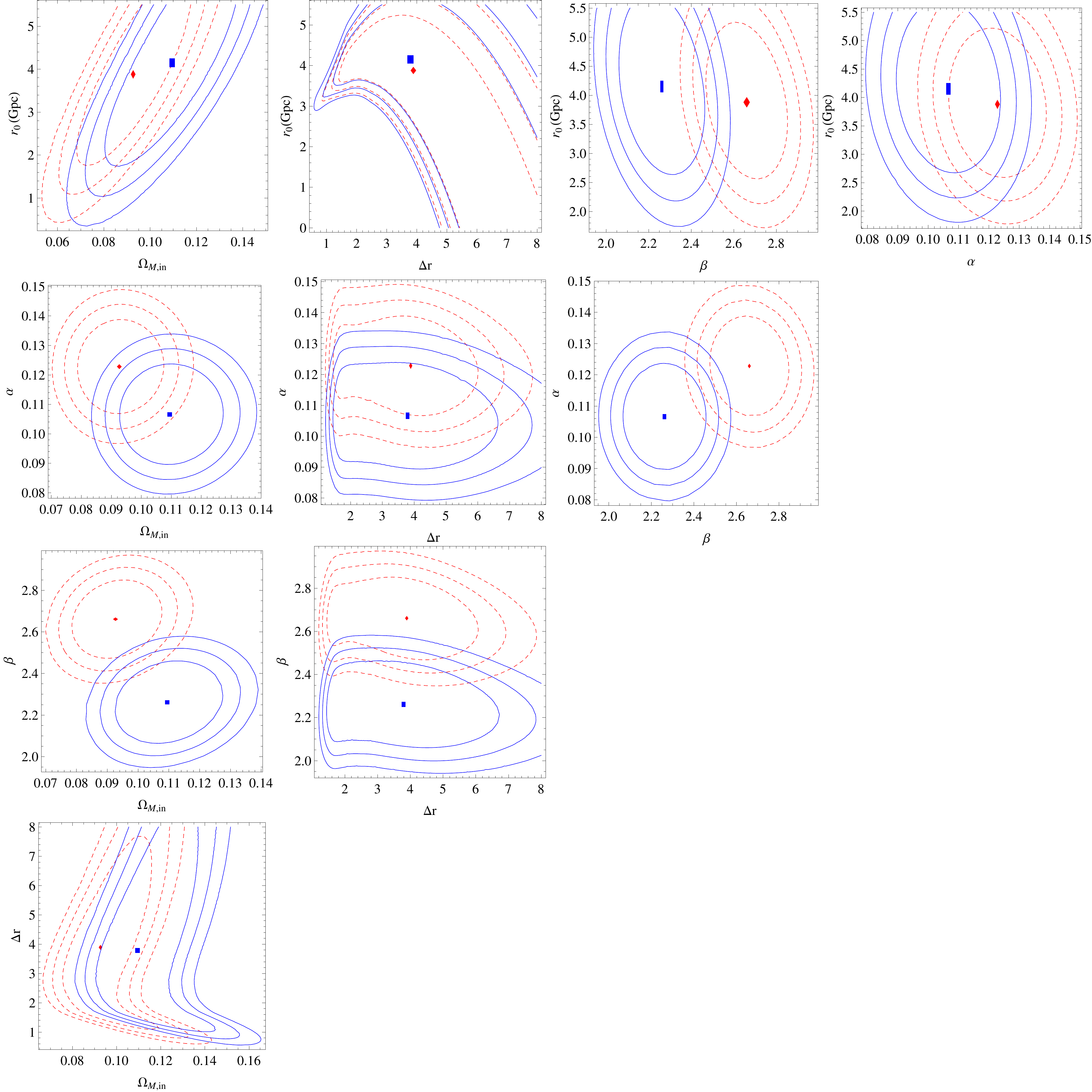}
\caption{Contour Level best fit parameters for $\chi^2$ (dashed red) and Likelihood (solid blue) approaches with the JLA sample only for the CGBH  model.  }  \label{fig:cgbh_sn}
\end{figure}

\begin{figure}
\includegraphics[width=12.cm,height=9.cm]{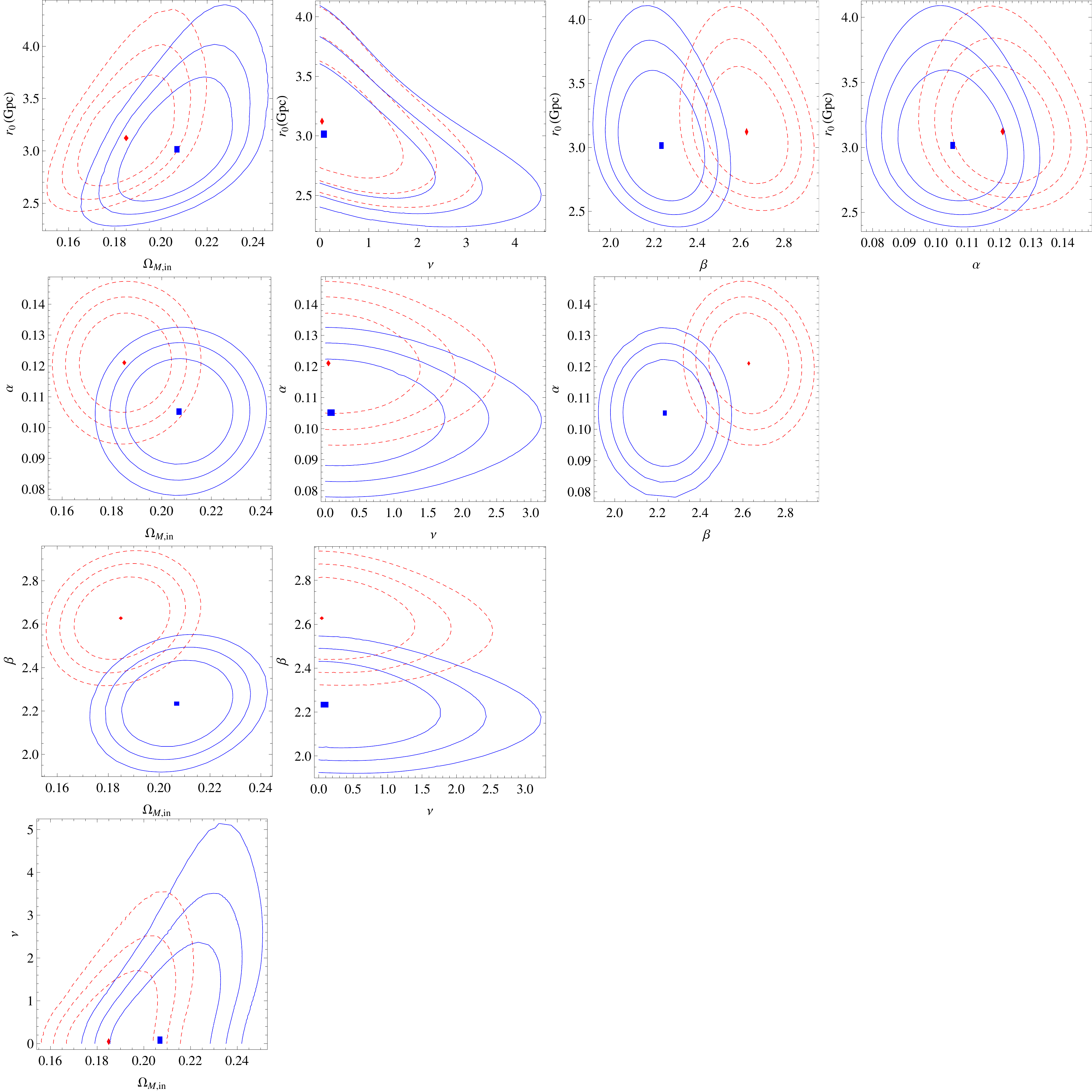}
\caption{Contour Level best fit parameters in the $\chi^2$ (dashed red) and $\mathcal{L}$ (solid blue) approaches with the JLA sample only for the C$\nu$-ln2 model.  }  \label{fig:ctrz_sn}
\end{figure}

\begin{table}
\centering
\begin{tabular}{|l| c c c c | c c c c | c c|}
\hline \hline
{\bf CGBH }  &  $\Omega _{M\tiny{\mbox{in}}}$  &    $r_0$  & $\Delta r$ & 100 $f_b$ &  $\alpha$  &  $\beta $ &  $M^{1}_B$  &  $\Delta _M$ &   $\chi ^{2}_{min}/\mbox{d.o.f.}$   &  $\sigma_\sube{int}$ \\
   \hline
  $\chi ^2_{JLA}$  &  0.09  & 3.88 & 3.89 & - & 0.12  &  2.66  &  -19.22  &  -0.04  &  713.61/733 & 0.02 \\
  \hline
  $\mathcal{L}_{JLA}$  &  0.11 & 4.15 & 3.79 & - & 0.11  &  2.26  &  -19.25  &  -0.03  &  - & 0.06 \\
  \hline
  $\chi ^2_{BAO}$ &  0.38  & 3.31 & 0.51 & 5.62 & -  &  -  &  -  &  -  &  1.03/3 & - \\
  \hline
  $\chi ^2_{JLA+BAO}$  &  0.123  & 3.16 & 0.73 & 15.3 & -  &  -  &  -  &  -  &  761.07/743 & 0.03 \\
  \hline
  $\mathcal{L}_{JLA+BAO}$  &  0.118  & 3.72 & 1.68 & 16.8 & -  &  -  &  -  &  -  &  - & 0.07 \\
\hline \hline
\end{tabular}
\caption{ Best fit parameters for the CGBH model in the $\chi ^2$ and Likelihood approaches.} \label{table2}
\end{table}
\begin{table}
\centering
\begin{tabular}{|l| c c c | c c c c | c c|}
\hline \hline
 {\bf Gaussian}  & \hspace{0.2cm} $\Omega _{M\tiny{\mbox{in}}}$  &  \hspace{0.2cm}  $r_0$  & \hspace{0.2cm} 100 $f_b$ \hspace{0.2cm} & \hspace{0.2cm} $\alpha$  &  $\beta $ &  $M^{1}_B$  &  $\Delta _M$ &   $\chi ^{2}_{min}/\mbox{d.o.f.}$   &  $\sigma_\sube{int}$ \\
   \hline
  $\chi ^2_{JLA}$  &  0.16  & 4.28 & - &  0.12  &  2.65  &  -19.27  &  -0.04  &  714.56/734 & 0.03 \\
  \hline
  $\mathcal{L}_{JLA}$  &  0.17 & 4.27 & - &  0.11  &  2.25 &  -19.30  &  -0.03  &  - & 0.07 \\
  \hline
  $\chi ^2_{BAO}$  &  0.26  & 7.39 & 18.9 &  -  &  -  &  -  &  -  &  1.09/4 & - \\
  \hline
  $\chi ^2_{JLA+BAO}$  &  0.17  & 5.08 & 16.31 &  -  &  -  &  -  &  -  &  741.52/744 & 0.02 \\
  \hline
  $\mathcal{L}_{JLA+BAO}$  &  0.19  & 5.04 & 14.49 & -  &  -  &  -  &  -  &  - & 0.07 \\
\hline \hline
\end{tabular}
\caption{Best fit parameters for the Gaussian model in the $\chi ^2$ and Likelihood approaches.} \label{table3}
\end{table}
\begin{table}
\centering
\begin{tabular}{|l| c c c c | c c c c | c c|}
\hline \hline
 {\bf C$\nu$ln-2}  &  $\Omega _{M\tiny{\mbox{in}}}$  &    $r_0$  & $\nu$ & 100 $f_b$ &  $\alpha$  &  $\beta $ &  $M^{1}_B$  &  $\Delta _M$ &   $\chi ^{2}_{min}/\mbox{d.o.f.}$   &  $\sigma_\sube{int}$ \\
   \hline
  $\chi ^2_{JLA}$  &  0.18  & 3.12 & 0.04 & - & 0.12  &  2.63  &  -19.29  &  -0.05  &  713.98/733 & 0.03 \\
  \hline
  $\mathcal{L}_{JLA}$  &  0.21 & 3.02 & 0.08 & - & 0.11  &  2.23 &  -19.33  &  -0.03  &  - & 0.07 \\
  \hline
  $\chi ^2_{BAO}$  &  0.26  & 5.23 & 0.58 & 19.54 & -  &  -  &  -  &  -  &  1.21/3 & - \\
  \hline
  $\chi ^2_{JLA+BAO}$  &  0.21  & 4.40 & 0.34 & 20.97 & -  &  -  &  -  &  -  &  742.59/743 & 0.14 \\
  \hline
  $\mathcal{L}_{JLA+BAO}$  &  0.22  & 3.54 & 0.03 & 11.65 & -  &  -  &  -  &  -  &  - & 0.07 \\
\hline \hline
\end{tabular}
\caption{Best fit parameters for the C$\nu$-ln2 model in the $\chi ^2$ and Likelihood approaches.} \label{table4}
\end{table}
\begin{table}
\centering
\begin{tabular}{|l| c c | c c c c | c c|}
\hline \hline
 {\bf $\Lambda$CDM}  & \hspace{0.2cm} $\Omega _{M}$   & \hspace{0.2cm} 100 $f_b$ \hspace{0.2cm} & \hspace{0.2cm} $\alpha$  &  $\beta $ &  $M^{1}_B$  &  $\Delta _M$ &   $\chi ^{2}_{min}/\mbox{d.o.f.}$   &  $\sigma_\sube{int}$ \\
   \hline
  $\chi ^2_{JLA}$  &  0.312  & - &  0.123  &  2.665  &  -19.022  &  -0.043  &  715.793/735 & 0.019 \\
  \hline
  $\mathcal{L}_{JLA}$  &  0.329 &  - &  0.107  &  2.265 &  -19.032  &  -0.028  &  - & 0.064 \\
  \hline
  $\chi ^2_{BAO}$  &  0.255  & 21.325 &  -  &  -  &  -  &  -  &  1.997/5 & - \\
  \hline
  $\chi ^2_{JLA+BAO}$  &  0.312  & 14.712  &  -  &  -  &  -  &  -  &  718.035/745 & 0.019 \\
  \hline
  $\mathcal{L}_{JLA+BAO}$  &  0.312  & 14.713 & -  &  -  &  -  &  -  &  - & 0.023 \\
\hline \hline
\end{tabular}
\caption{Best fit parameters for the $\Lambda$CDM model in the $\chi ^2$ and Likelihood approaches.} \label{tableLCDM}
\end{table}
The contour levels for each profile using only SNIa, i.e. only the JLA sample, are depicted in figures~\ref{fig:gaussian_sn},~\ref{fig:cgbh_sn} and~\ref{fig:ctrz_sn}. In these figures, the dashed contours indicate the $1\sigma$, $2\sigma$ and $3\sigma$ confidence regions for the $\chi^2$ approach, while the solid lines show the contours for the Likelihood approach. Figure~\ref{fig:gaussian_sn} displays the Gaussian profile while the CGBH and C$\nu$-ln2 profiles are shown respectively, in fig.~\ref{fig:cgbh_sn} and~\ref{fig:ctrz_sn}. Note that the contour levels for the $\chi^2$ and likelihood approaches are similar in shape and area. However, there is an interesting difference from the results of Ref.~\cite{Ribamar}. Here there is a significant bias in the best fit values and the likelihood contours are slightly bigger than the $\chi^2$.

\begin{figure}
\centering
\includegraphics[width=4cm,height=3cm]{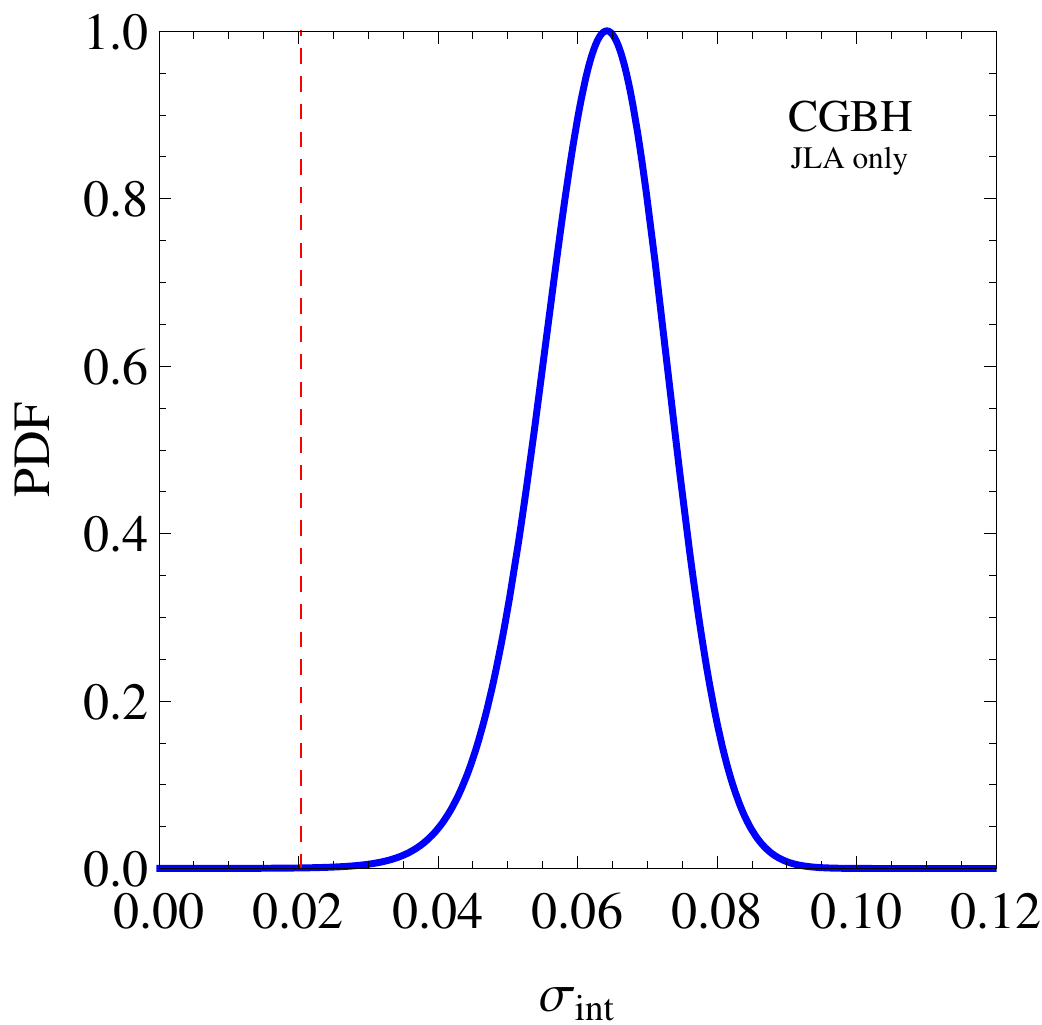}
\includegraphics[width=4cm,height=3cm]{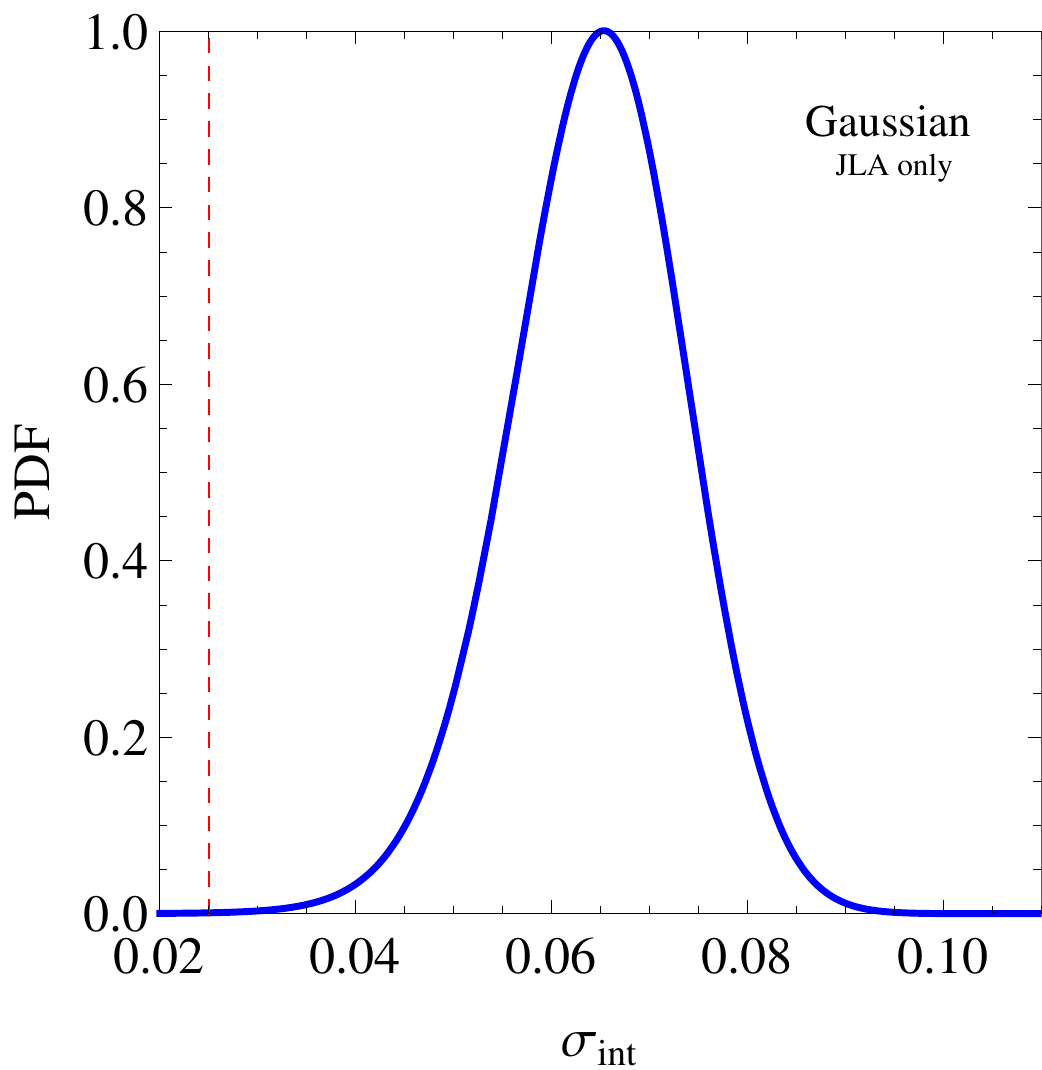}
\includegraphics[width=4cm,height=3cm]{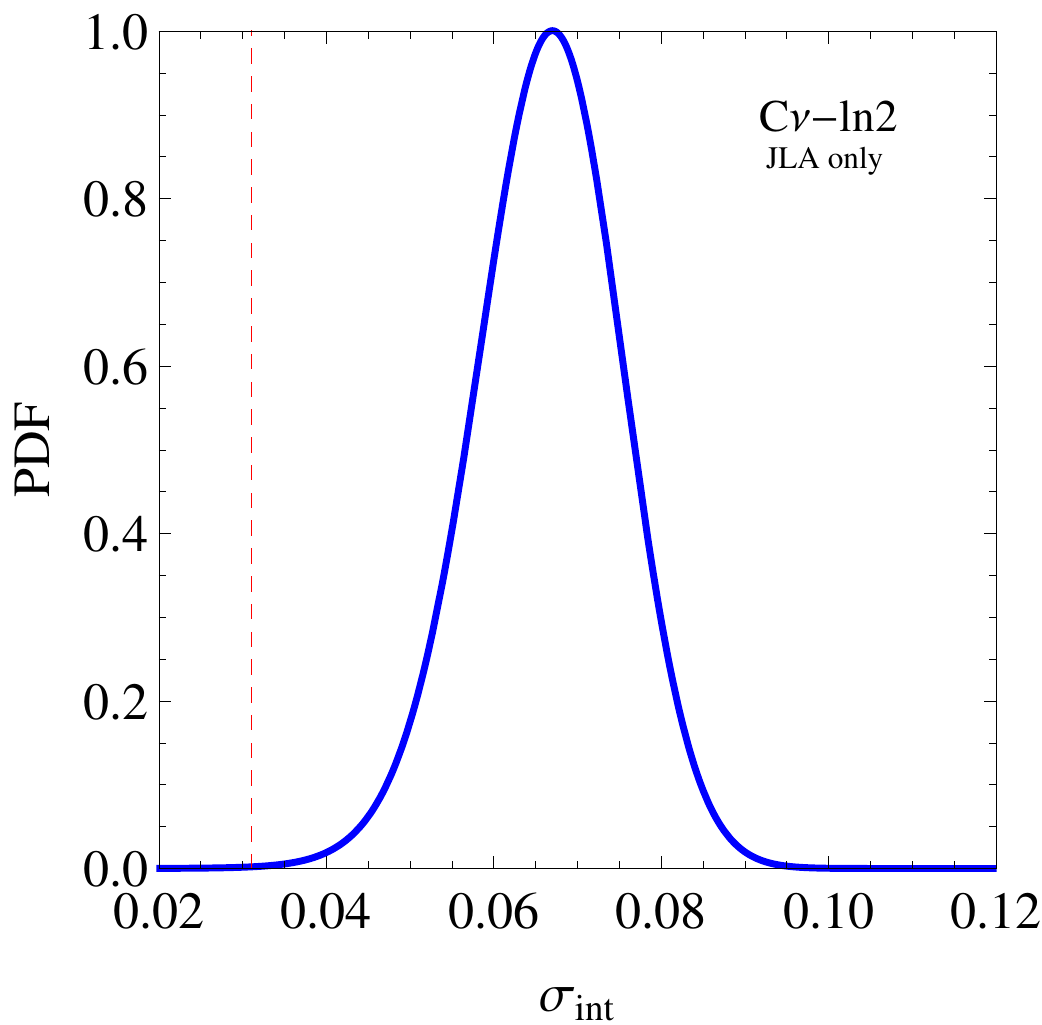}
\caption{PDF for $\sigma_\sube{int}$ in the Likelihood approach for the CGBH (left panel), Gaussian (middle panel) and C$\nu$-ln2 (right panel) models with the JLA only. The red dashed line indicate the values obtained in the $\chi^2$ approach.}\label{fig:sigmaint_pdf_sn}
\end{figure}

\begin{figure}
\centering
\includegraphics[width=4cm,height=3cm]{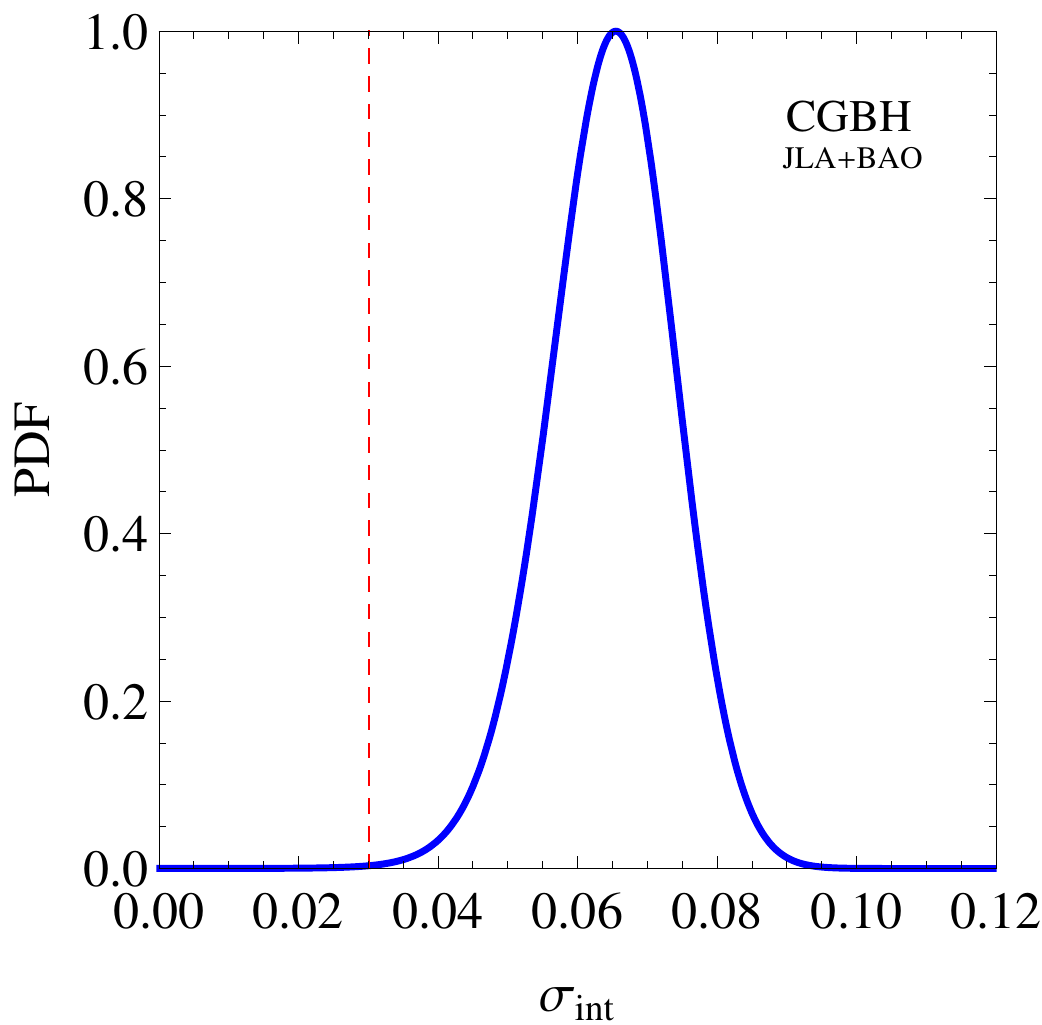}
\includegraphics[width=4cm,height=3cm]{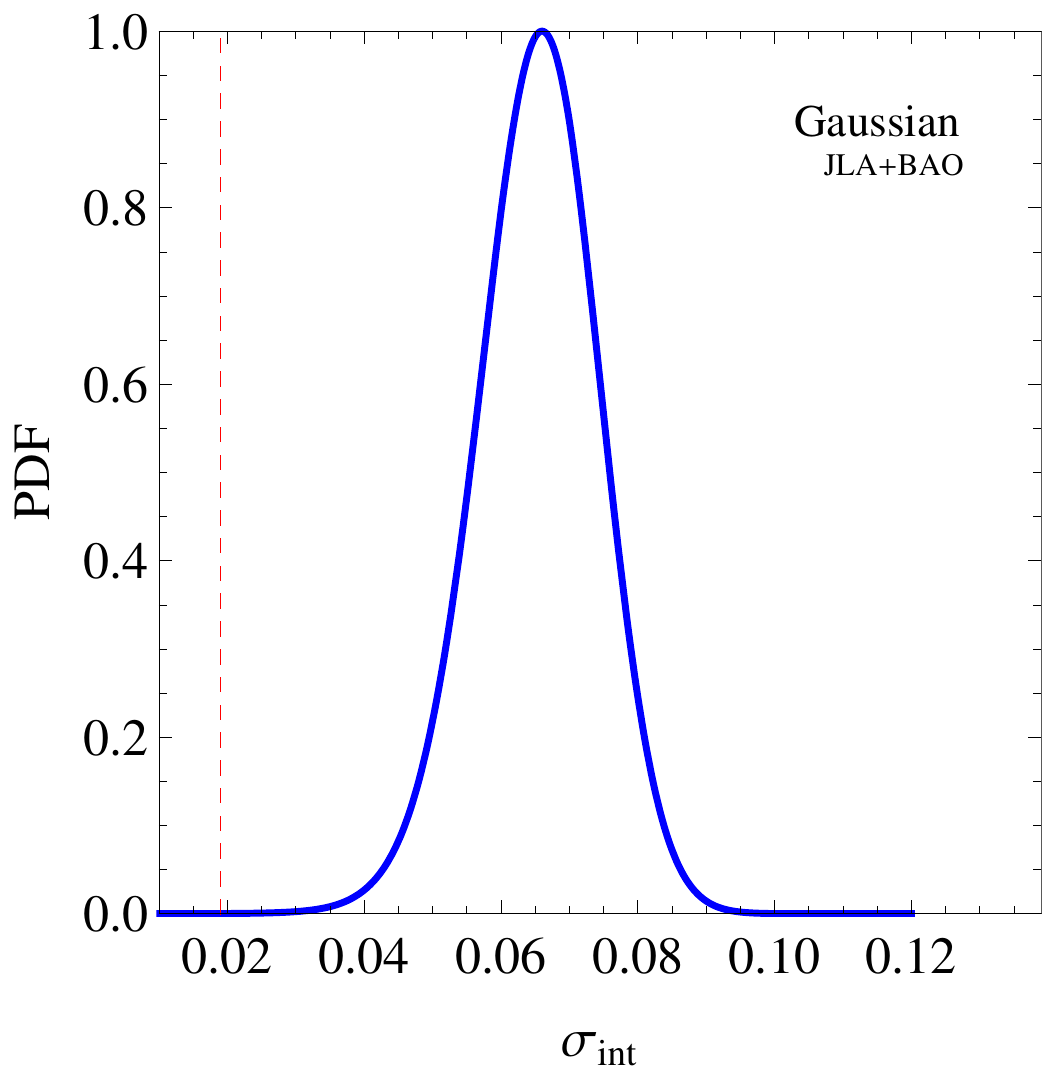}
\includegraphics[width=4cm,height=3cm]{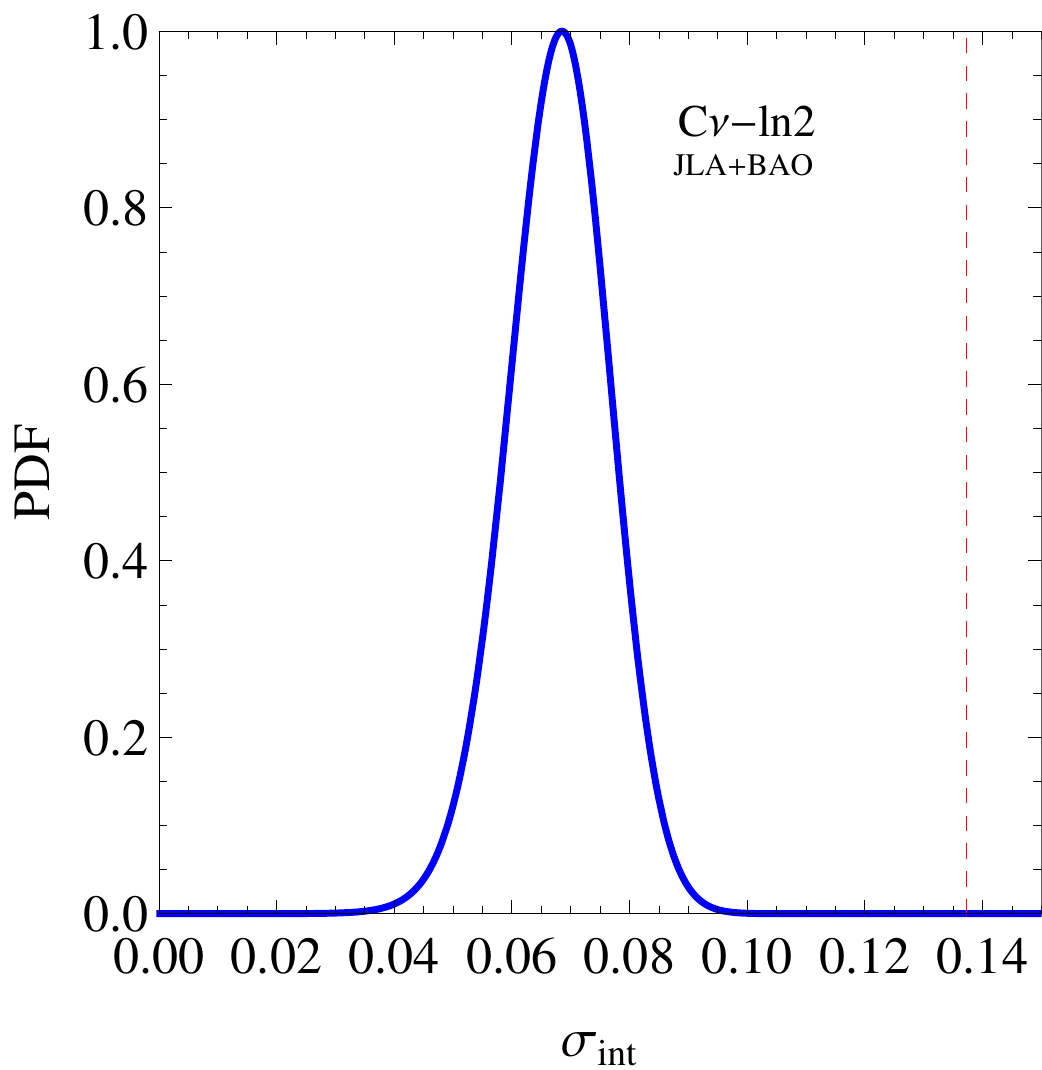}
\caption{PDF for $\sigma_\sube{int}$ in the Likelihood approach for the CGBH (left panel), Gaussian (middle panel) and C$\nu$-ln2 (right panel) models with the combined JLA+BAO. The red dashed line indicate the values obtained in the $\chi^2$ approach.}\label{fig:sigmaint_pdf_comb}
\end{figure}

Figs.~\ref{fig:sigmaint_pdf_sn} and~\ref{fig:sigmaint_pdf_comb} display the 1D probability distributions function (PDF) of the $\sigma_{int}$ parameter for all models. Fig.~\ref{fig:sigmaint_pdf_sn} considers only the JLA sample while fig.~\ref{fig:sigmaint_pdf_comb} depicts the combined analysis of JLA+BAO. The dashed straight line in each panel indicates the value obtained from the $\chi^2$ approach. Note that in all cases the $\chi^2$ values are excluded with respect to the likelihood approach. This can also be seen through the contour levels. The two approaches give disjoint results at $1 \sigma$ in the planes ($\beta$, $r_0$), ($\beta$,$\alpha$), ($\Delta r$, $\beta$), ($\nu$, $\beta$) and ($\Omega _{M\tiny{\mbox{in}}}$, $\beta$). Thus, our analysis exhibits a concrete example of the issues and criticisms related to the $\chi^2$ approach discussed in Ref.~\cite{Ribamar}.

\begin{figure}
\centering
\includegraphics[width=6.cm,height=6cm]{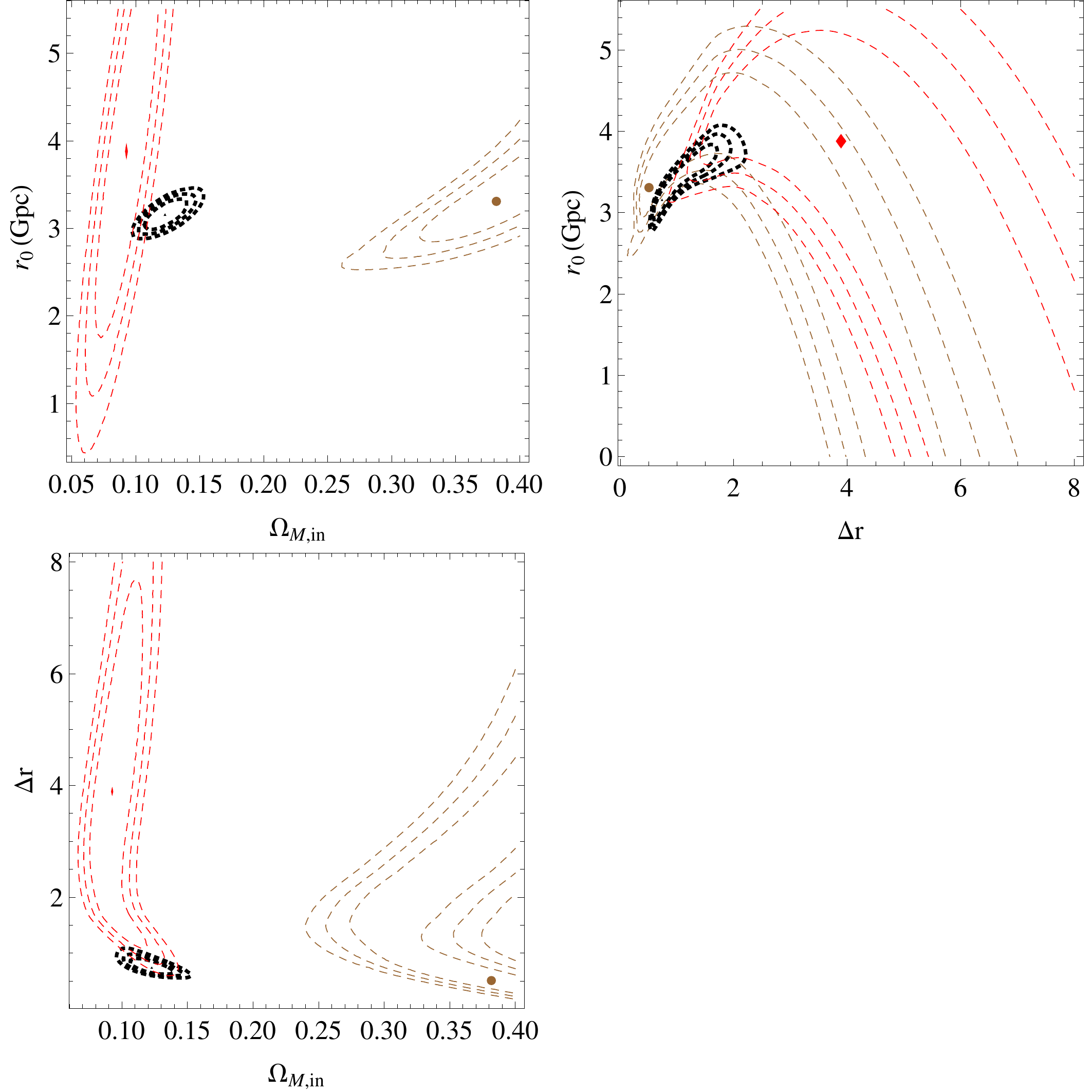}
\includegraphics[width=6.cm,height=6cm]{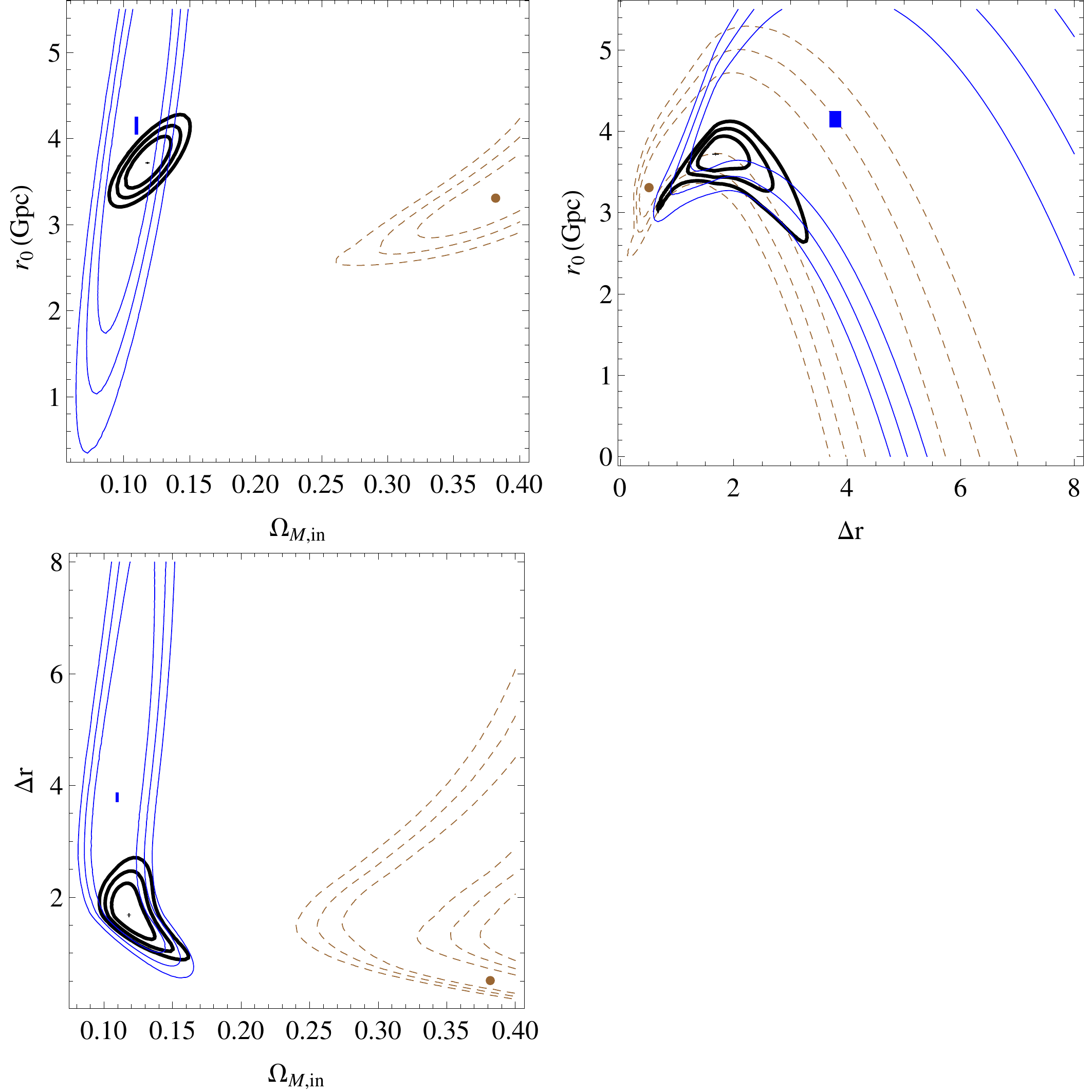}
\caption{$1\sigma$, $2\sigma$ and $3\sigma$ contour Level best fit parameters for the CGBH model in the $\chi ^2$ approach (three left panel), with the JLA only (dashed red) and the combined JLA+BAO (dotted black), and in the Likelihood approach (three right panel), with the JLA only (solid blue) and the combined JLA+BAO (solid black). The dashed brown are the contours for the BAO only.}\label{fig:contoursJLABAOcgbh}
\end{figure}

\begin{figure}
\centering
\includegraphics[width=4cm,height=3cm]{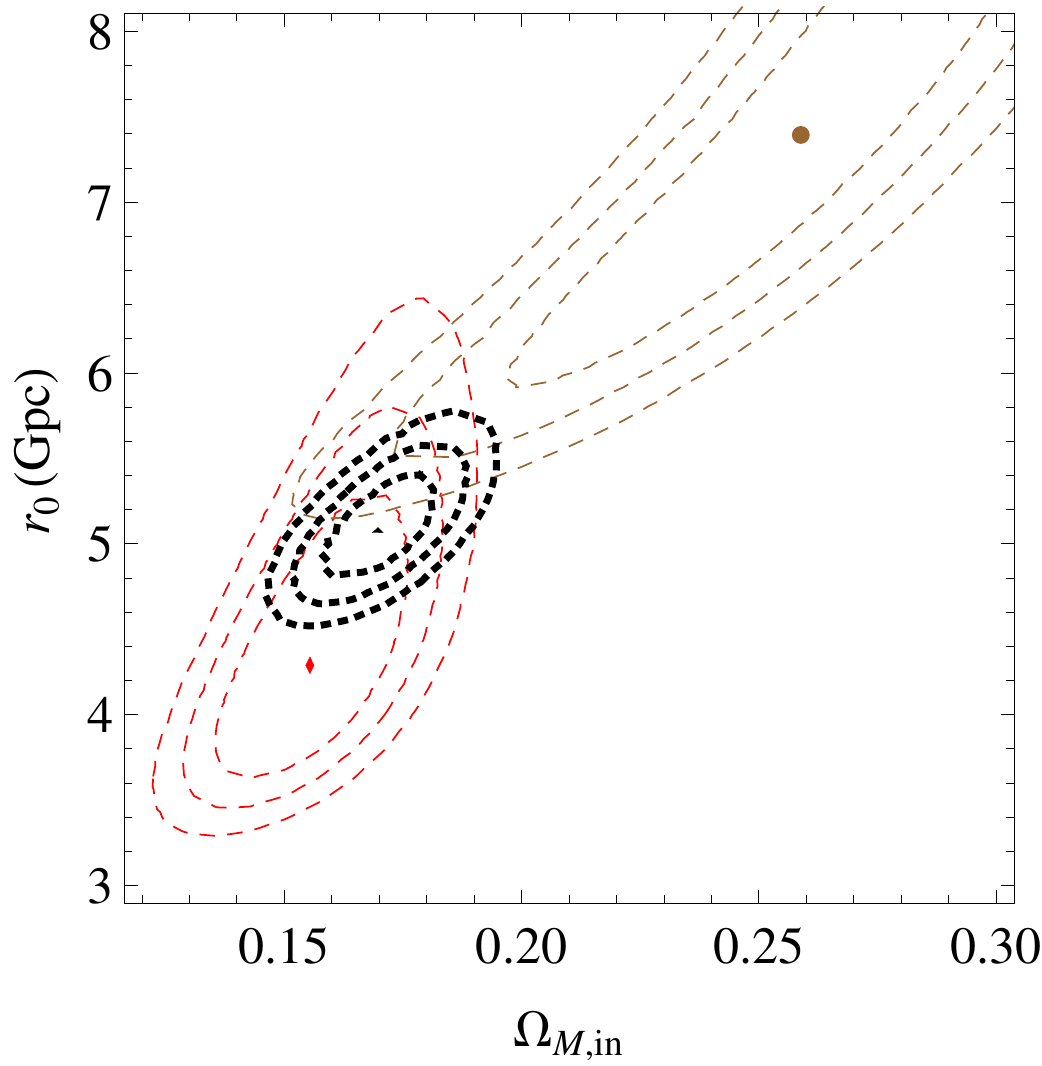}
\includegraphics[width=4cm,height=3cm]{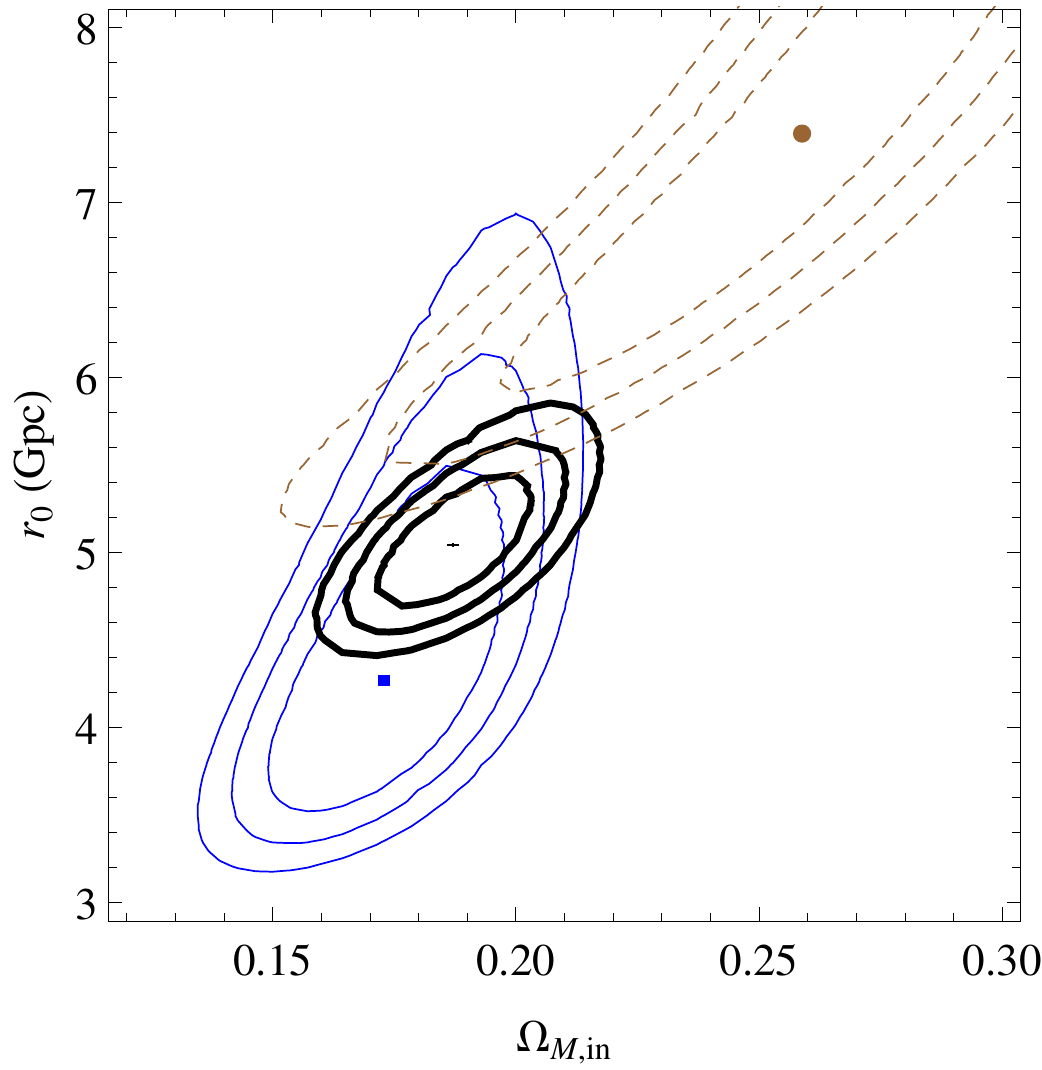}
\caption{$1\sigma$, $2\sigma$ and $3\sigma$ contour Level best fit parameters for the Gaussian model in the $\chi ^2$ approach (left panel), with the JLA only (dashed red) and the combined JLA+BAO (dotted black), and in the Likelihood approach (right panel), with the JLA only (solid blue) and the combined JLA+BAO (solid black). The dashed brown are the contours for the BAO only..}\label{fig:contoursJLABAOgaussian}
\end{figure}

\begin{figure}
\centering
\includegraphics[width=6cm,height=6cm]{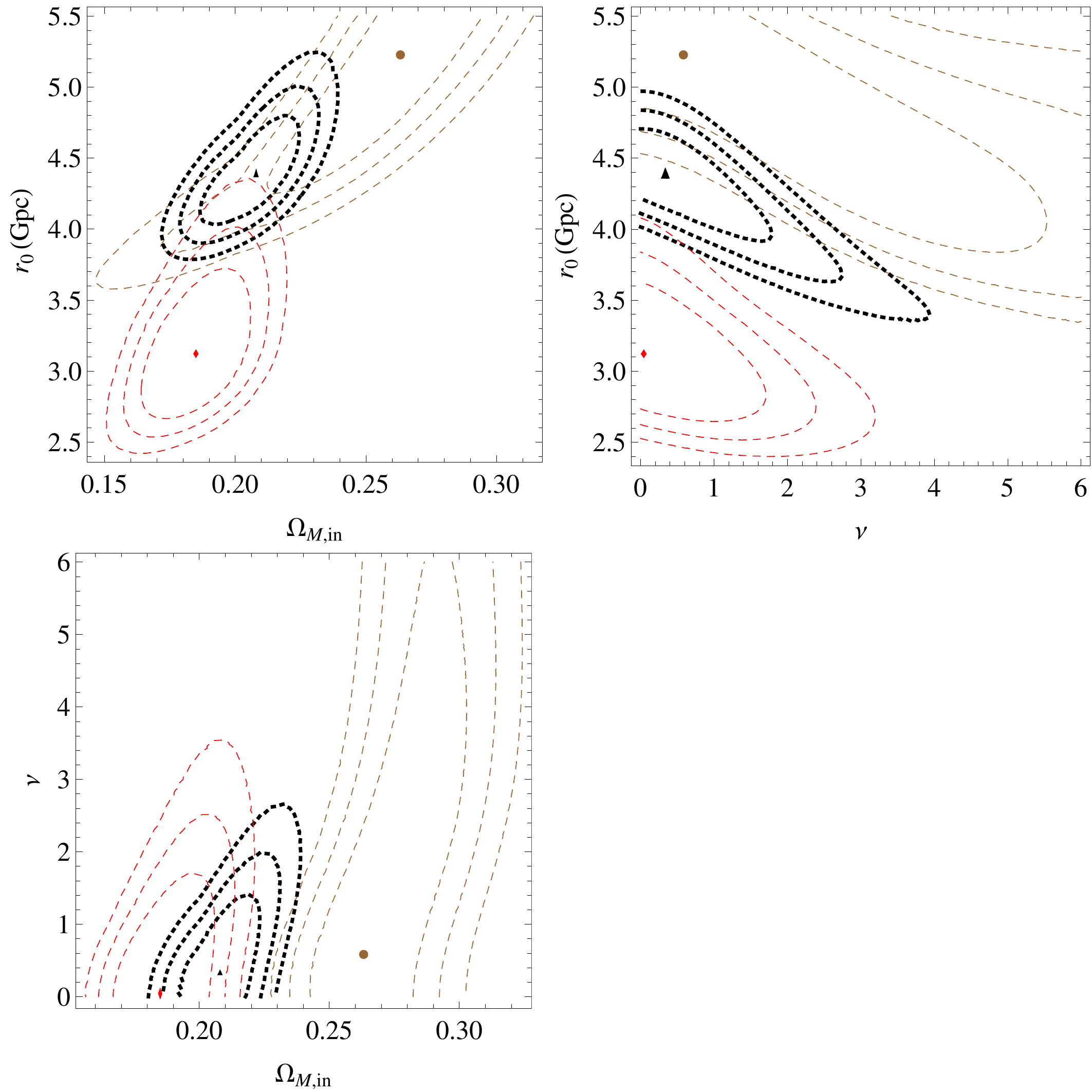}
\includegraphics[width=6cm,height=6cm]{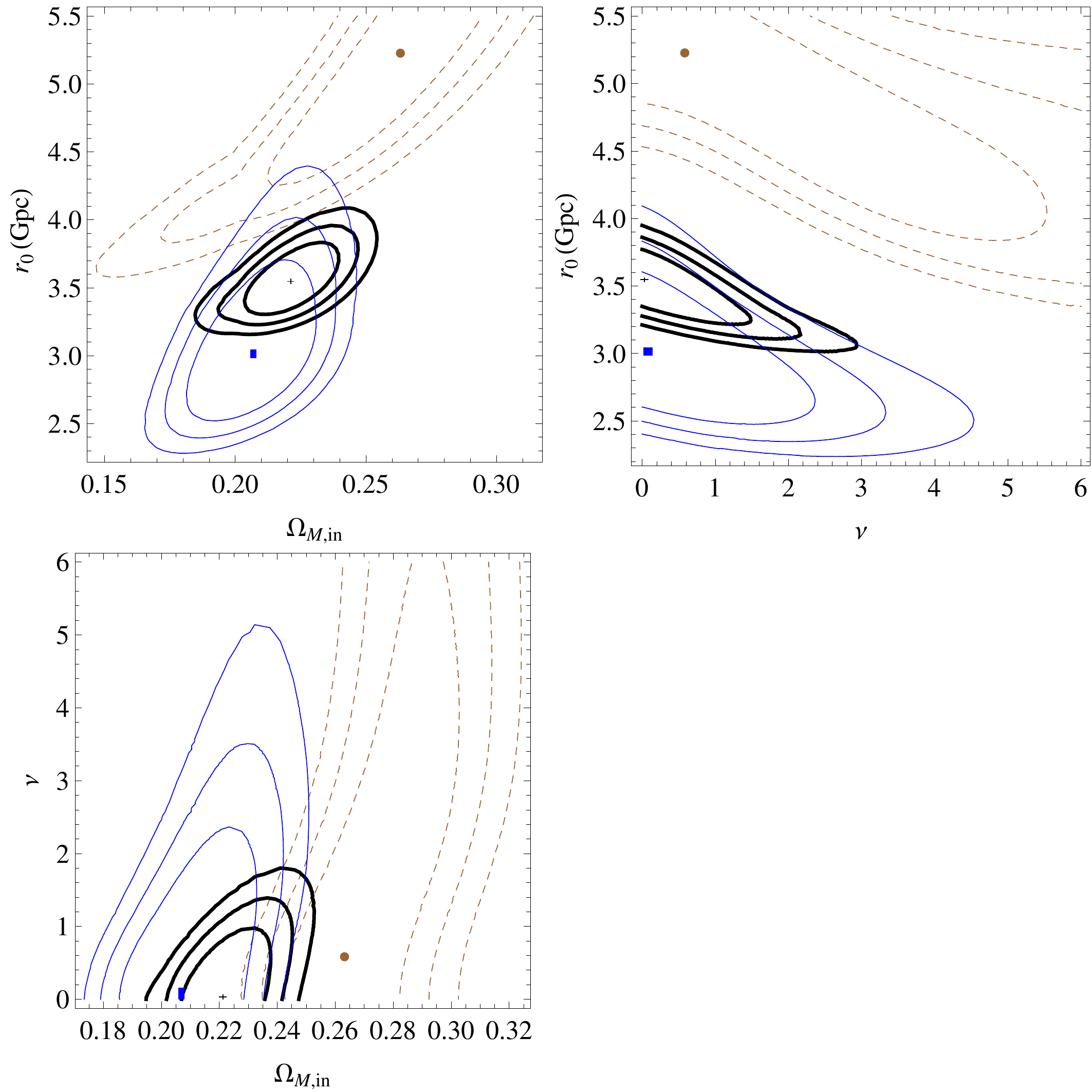}
\caption{$1\sigma$, $2\sigma$ and $3\sigma$ contour Level best fit parameters for the C$\nu$-ln2 model in the $\chi ^2$ approach (three left panel), with the JLA only (dashed red) and the combined JLA+BAO (dotted black), and in the Likelihood approach (three right panel), with the JLA only (solid blue) and the combined JLA+BAO (solid black). The dashed brown are the contours for the BAO only.}\label{fig:contoursJLABAOcnln2}
\end{figure}

In Section~\ref{subsec:BAOinLTB} we introduced the parameter $f_b$ and described how one can use standard rulers (BAO) within LTB scenarios. Independently of the statistical approach used ($\chi^2$ or Likelihood), it appears \textcolor{blue}{a tension} between BAO and SNIa best fit results (see the planes ($\Omega _{M\tiny{\mbox{in}}}$, $r_0$) and ($\Omega _{M\tiny{\mbox{in}}}$, $\Delta r$) in fig.~\ref{fig:contoursJLABAOcgbh} and the planes ($\Omega_{M\sube{in}}$, $\nu$) and ($\nu$, $r_0$) in fig.~\ref{fig:contoursJLABAOcnln2}). It has been argued in the literature (see Ref.~\cite{zumalacarregui}) that this discrepancy comes from the evolution of a non-zero shear in LTB models. Indeed, the spatial dependence and the difference in the evolution of the two LTB expansion rates ($H_{\parallel}$ and $H_{\bot}$) works differently in fitting BAO and SNIa data. In particular, the low value of $\Omega _{M\tiny{\mbox{in}}}$ needed to fit the SNIa data increases the expansion rate that end up over-stretching the BAO scale near the center. In this sense, the tension observed in our analysis has a pure geometrical origin associated with the LTB dynamics. 
Figures~\ref{fig:contoursJLABAOcgbh}, \ref{fig:contoursJLABAOgaussian} and~\ref{fig:contoursJLABAOcnln2}  show that using only BAO to fit the parameters disagrees with the best fit of using only SNIa at least at $3\sigma$ of confidence level.

In the case of the CGBH profile model, fig.~\ref{fig:contoursJLABAOcgbh} shows that the BAO data favors a denser void (higher matter density) as compared to the SNIa data. For the Gaussian model, fig.~\ref{fig:contoursJLABAOgaussian} indicates that BAO favors not only higher values of matter density inside the void but also bigger voids, i.e. higher values of parameter $r_0$. In the case of the C$\nu$-ln2, there is still the same tension between BAO and SNIa best fit values yet in smaller extend as compared to the CGBH model.

There are several ways of performing model comparison. For instance, the (corrected) Akaike Information Criterion\footnote{We have reintroduced the term $N \ln (2\pi)$, in AIC and BIC, in order to obtain the $-2 \ln \hat{L}$ value, where $\hat{L}$ is the maximum likelihood function.} (AIC)~\cite{akaike} proposes to compare different models through a quantity defined as

\begin{eqnarray}
AIC &=& \mathcal{L}_{min}+ N \ln (2\pi) + 2 k + \frac{2 k (k -1)}{N - k - 1} \ , 
\end{eqnarray}
where $k$ is the number of free parameters and $N$ is the number of data points. Another possibility is the Bayesian Information Criterion (BIC)~\cite{bic} which uses the quantity
\begin{eqnarray}
BIC &=& \mathcal{L}_{min} + N \ln (2\pi) + k \ln N \ .
\end{eqnarray}
A model is viewed as favored by the data when a lower AIC or BIC value is obtained. Note that their difference comes from the last two terms in AIC and the last one in BIC. Thus, for large amount of data points ($N\gg 1$) BIC is a little more sensitive to an increment of parameter than AIC. Indeed, by an increment $k\rightarrow k+\delta_k$, AIC changes by an amount $2\delta_k\left[1+(\delta_k + 2k-1)/(N-\delta_k - k -1)\right] + \mathcal{O}(N^{-2})$ while BIC is proportional to $\delta_k \ln N$.

Table~\ref{table5} summarize the results of each profile and the $\Lambda$CDM model for SNIa and SNIa+BAO using the AIC and BIC information criterions. In both criterions the $\Lambda$CDM is favored with respect to the LTB models.    
Note that for the combined analysis SNIa+BAO, the $\Lambda$CDM is notoriously more favored.
\begin{table}
\centering
\begin{tabular}{|l| c c | c c |}
\hline \hline
  & \multicolumn{2}{|c|}{JLA} &  \multicolumn{2}{|c|}{JLA + BAO}  \\
   \hline
 {\bf Model}  & AIC & BIC &  AIC  & BIC  \\
   \hline
  CGBH   & -621.118 & -584.418 & -602.296  &  -579.318 \\
   \hline
  Gaussian   & -618.736 & -586.604 & -622.458  &  -604.064 \\
   \hline
  C$\nu$-ln2   & -610.448 & -573.748 & -607.356  &  -584.378 \\
  \hline
  $\Lambda$CDM   & -623.689 & -596.131 & -666.025  &  -652.221 \\
\hline \hline
\end{tabular}
\caption{Comparison of the information criterion for the best-fit parametrization of the three void models and the $\Lambda$CDM model using JLA and JLA+BAO data.} \label{table5}
\end{table}
In order to obtain the Bayes factor $B_{ij}$, we can use a rough approximation~\cite{kass} that is worthy as $N \rightarrow \infty $. In this limit it can be shown that
\begin{eqnarray}\label{lnBij}
\frac{BIC[i]-BIC[j]+ 2 \ln B_{ij}}{2 \ln B_{ij}} \rightarrow 0 \ ,
\end{eqnarray}
where $BIC[j]$ denote BIC for model $j$. This relation is know as the Schwarz criterion and does not give the precise value of $B_{ij}$ but it is easer to manage and does not require evaluation of prior densities trough the use of the Maximum Likelihood Estimator to the parameters on each model. Of course, the relative error of this approximation is higher (of order $\mathcal{O}(1)$) and are not getting the correct value of the Bayes factor. But, keeping in mind the rough interpretation of the Bayes factor on the logarithmic scale suggested in sec 3.2 of~\cite{kass}, the relation~\eqref{lnBij} shows that in larges samples its should provide a reasonable indication of the evidence. Our results are summarized in Table~\ref{table6}, where we denote respectively the $\Lambda$CDM, Gaussian, CGBH and C$\nu$-ln2 cases as model 1, 2, 3 and 4.  It is worth noting that for the analysis using only SNIa data the evidence against the Gaussian model is strong ($6 \leq 2 \ln B_{ij} \leq 10$) in contrast with both CGBH and C$\nu$-ln2 models which indicate very strong evidence ($2 \ln B_{ij} > 10$) to the $\Lambda$CDM model. For the JLA+BAO combined analysis the evidence of the $\Lambda$CDM is very strong against the three LTB models.

\begin{table}
\centering
\begin{tabular}{| l | c  |  c |}
\hline \hline
   &  JLA  &    JLA + BAO \\
   \hline
 $2 \ln B_{12}$  &  9.527  & 48.157  \\
   \hline
  $2 \ln B_{13}$  &  11.713  & 72.903   \\
  \hline
  $2 \ln B_{14}$  &  22.383  & 67.843   \\
\hline \hline
\end{tabular}
\caption{Bayes factor for JLA and JLA+BAO, considering the $\Lambda$CDM, Gaussian, CGBH and C$\nu$-ln2  respectively as model 1, 2, 3  and 4.} \label{table6}
\end{table}

\section{Conclusions}\label{sec:Concl}

In this paper we have presented exact inhomogeneous LTB models that can suppress the need of dark energy to explain the present acceleration of the universe. Within this scenario, the observational data indicate that we live in a large inhomogeneous void of the order of few Gigaparsec. In particular, we considered three different profiles of void LTB universe models. Both the CGBH and C$\nu$-ln2 profiles have three cosmological parameters unlike the Gaussian profile that has only two parameters. In order to include the BAO analysis, it has been added an extra parameter $f_b$ that represent the Baryon fraction over the total matter content. In addition, there is also the supernovae nuisance parameters in the distance estimate which are the same in all cases. 

We performed the best fit analysis and constrained the space of parameters for the inhomogeneous models. This analysis was carried out using BAO and SNIa data separately and also the JLA+BAO combined analysis. We have been careful to calibrate SNIa data specifically for the LTB dynamics. The dependence on the background dynamics does not change significantly from a FLRW background but this calibration is necessary and should be checked every time. 

We have also tested the validity of the $\chi^2$ minimization compared to the complete Likelihood approach. Our results corroborate with Ref.~\cite{Ribamar} presenting a specific example where these two approaches are not equivalent. The supernovae parameters ($\alpha$, $\beta$, $M^ 1_{B}$ and $\Delta_M$) have similar results for all cases, but presents bias in both approaches. On the other hand, the cosmological parameters ($\Omega_{\scriptsize{M \sube{in}}}$, $\Delta r$, $\nu$ and $r_0$) have considerable deviation regardless of the method. For example, the C$\nu$-ln2 model presents higher matter density $\Omega_{\scriptsize{M \sube{in}}}$ than the CGBH and Gaussian models. The contour levels for the $\chi^2$ and Likelihood approaches display similar shapes and areas but significant bias in the best fit values with the likelihood contours slightly bigger than the $\chi^2$.

Figures~\ref{fig:sigmaint_pdf_sn} and~\ref{fig:sigmaint_pdf_comb} show the 1D probability distribution function (PDF) of the $\sigma_\sube{int}$ parameter for the three models, considering respectively only the JLA sample and the JLA+BAO combined analysis. These results also agree with the issues and criticisms related to the $\chi^2$ analysis.

We found a tension between the confidence contours coming separately from SNIa and BAO data. This discrepancy stems from the behavior of the LTB radial and transverse expansion rates, which works differently for BAO and SNIa. Indeed, the low value of  $\Omega_{\scriptsize{M \sube{in}}}$ needed to fit the SNIa data increases the expansion rate that consequently stretch the BAO scale near the center. This discrepancy was seen at more than 3$\sigma$ in Figure~\ref{fig:contoursJLABAOcgbh}, for CGBH, and at lower confidence level for Gaussian and C$\nu$-ln2 models (Figs. \ref{fig:contoursJLABAOgaussian} and \ref{fig:contoursJLABAOcnln2}). 

Finally, we have analyzed the AIC and BIC information criterions in order to evaluate the best model. The Gaussian model is slightly favored in comparison with the CGBH and C$\nu$-ln2 models. But still the $\Lambda$CDM is the best favored. The main difference in the profiles is the number of parameter where the Gaussian profile has one less than the others. This is an advantage in the information criterion and given the proximity of the result it is arguably the reason for it to exceed the other models. We have also calculated an approximate Bayes factor, which when using only SNIa data indicates a strong evidence to the $\Lambda$CDM model against Gaussian model in contrast with both the CGBH and C$\nu$-ln2 models that suggest a very strong evidence to the $\Lambda$CDM model. For the combined BAO+SNIa analysis, the $\Lambda$CDM model have a very strong evidence against the three LTB models.

\section*{Acknowledgements}

We acknowledge the Conselho Nacional de Desenvolvimento Cient\'ifico e Tecnol\'ogico (CNPq - Brazil) and the Coordena\c{c}\~ao de Aperfei\c{c}oamento de Pessoal de N\'ivel Superior (CAPES) for financial support.


\end{document}